\documentclass[aps,prd,preprint,nofootinbib,a4paper,11pt,superscriptaddress]{revtex4-1}

\usepackage[T2A,T1]{fontenc}
\usepackage[centertags]{amsmath}
\usepackage{amsfonts}
\usepackage{amssymb}
\usepackage[sort&compress]{natbib}% упорядочивает ссылки
\usepackage[hypertex]{hyperref}%hypertex
\usepackage[russian,english]{babel}
\usepackage[dvips]{graphicx}% рисунки

\DeclareMathOperator{\re}{Re}
\DeclareMathOperator{\im}{Im}
\DeclareMathOperator{\Li}{Li}
\DeclareMathOperator{\Sp}{Sp}
\DeclareMathOperator{\res}{res}
\DeclareMathOperator{\sgn}{sgn}

\newcommand{\Z}{\mathbb{Z}}
\newcommand{\spx}{\mathbf{x}}
\newcommand{\spp}{\mathbf{p}}

\newcommand{\e}{\varepsilon}
\newcommand{\vf}{\varphi}
\newcommand{\s}{\sigma}

\newcommand{\al}{\alpha}
\newcommand{\ga}{\gamma}
\newcommand{\Ga}{\Gamma}
\newcommand{\de}{\delta}
\newcommand{\De}{\Delta}
\newcommand{\ka}{\varkappa}
\newcommand{\la}{\lambda}

\newcommand{\ups}{\upsilon}

\begin{document}

\selectlanguage{english}

\title{One-loop omega-potential of quantum fields with ellipsoid constant-energy surface dispersion law}

\date{\today}

\author{P.O. Kazinski}
\email[E-mail:]{kpo@phys.tsu.ru}
\affiliation{Physics Faculty, Tomsk State University, Tomsk, 634050 Russia}
\affiliation{Institute of Monitoring of Climatic and Ecological Systems, SB RAS, Tomsk, 634055 Russia}
\author{M.A. Shipulya}
\email[E-mail:]{sma@phys.tsu.ru}
\affiliation{Physics Faculty, Tomsk State University, Tomsk, 634050 Russia}

\begin{abstract}

Rapidly convergent expansions of a one-loop contribution to the partition function of quantum fields with ellipsoid constant-energy surface dispersion law are derived. The omega-potential is naturally decomposed into three parts: the quasiclassical contribution, the contribution from the branch cut of the dispersion law, and the oscillating part. The low- and high-temperature expansions of the quasiclassical part are obtained. An explicit expression and a relation of the contribution from the cut with the Casimir term and vacuum energy are established. The oscillating part is represented in the form of the Chowla-Selberg expansion for the Epstein zeta function. Various resummations of this expansion are considered. The developed general procedure is applied to two models: massless particles in a box both at zero and non-zero chemical potential; electrons in a thin metal film. The rapidly convergent expansions of the partition function and average particle number are obtained for these models. In particular, the oscillations of the chemical potential of conduction electrons in graphene and a thin metal film due to a variation of sizes of the crystal are described.

\end{abstract}


\maketitle

\section{Introduction}

Evaluation of one-loop corrections is a cornerstone mean allowing to describe the quantum corrections to physical properties of many-particle systems. In many cases, the one-loop corrections together with their renormalization group improvements provide the only attainable information about quantum features of the system which can be analyzed analytically. An identity of the formal expressions for these corrections makes to use the same computational techniques in different fields of physics from astrophysics and cosmology with characteristic energy scales up to hundreds of GeVs down to condensed matter physics with energies of the order of a few eVs and lesser. The calculations of the one-loop corrections are confronted with certain mathematical difficulties which are common for all these branches of physics. The problem becomes extremely complicated when we investigate quantum fields with imposed non-trivial boundary conditions and/or interacting with background fields of a complex structure.

There is a well elaborated quasiclassical approach known as the heat kernel expansion for how to calculate the one-loop corrections both at zero (see for review \cite{VasilHeatKer}) and finite temperatures and chemical potentials \cite{DowKen,KirstHKE} for almost any background fields and boundary conditions. Nevertheless, it possesses some weaknesses. This method is mainly developed for particles with a relativistic dispersion law (see, however, \cite{Gilkey}). Besides, and what even is more important, it does not catch exponentially suppressed terms which are substantial and observable at certain values of external fields. In condense matter physics, the classical example of these contributions provide the oscillating terms in the $\Omega$-potential of conduction electrons in metals. It was apparently Landau \cite{LandOsc,Shoenb} who first observed that an applied magnetic field may result in the oscillations of the partition function of conduction electrons. These oscillation were experimentally discovered by Schubnikov and de Haas \cite{SchubdeHa} in conductivity and by de Haas and van Alphen \cite{deHavAlp} in magnetic susceptibility. Later on, a quasiclassical theory of these oscillations was developed for the Fermi liquid quasiparticles with an arbitrary dispersion law \cite{OnsagOsc,Shoenb,LifshKos,Nedore,Azbel,Kulik,LifshKag}.

In this paper, we are going to consider the free quantum fields subjected to certain boundary conditions. The interaction with an environment and the self-interaction are taken into account only in the form of the dispersion law which we choose to be
\begin{equation}\label{disper law}
    E(p)=\omega(g^{ij}(p_i+a_i)(p_j+a_j)+m^2),
\end{equation}
where $g_{ij}$, $a_i$, and $m$ are some constant quantities, $g_{ij}$ being symmetric and positive definite, and $\omega(x)$ is some smooth function increasing at a large argument. Henceforth, the Einstein summation rule is assumed and indices are risen and lowered by the ``metric'' $g^{ij}$ and its inverse. This metric is related to the effective mass tensor,
\begin{equation}
    m_{ij}:=\left(\frac{\partial^2 E}{\partial p_i\partial p_j}\right)^{-1}_{p_i=-a_i}=\frac{g_{ij}}{\omega'(m^2)},
\end{equation}
when the latter is defined. Such a problem definition is typical for the Fermi liquid theory of conduction electrons in metals where the dispersion laws take exotic forms largely deviating from the familiar non-relativistic and relativistic ones (see, e.g., \cite{LifshKag}). Assuming relation \eqref{disper law}, we consider the simplest modification of the dispersion law such that its constant-energy surface, and, in particular, the Fermi surface, has an ellipsoid form. Of course, the form of the dispersion law \eqref{disper law} chosen by us is not accidental. Only in this case, we can reduce the problem of analytic evaluation of the partition function to a one-dimensional one. This, in turn, admits us to investigate characteristic features of the $\Omega$-potential in details analytically going beyond the quasiclassical approximation of the general theory of oscillations \cite{LifshKag}. Our aim is to derive the rapidly convergent expansions with a controllable error of the one-loop partition function for particles with the dispersion law \eqref{disper law}.

In calculating the $\Omega$-potential, we shall see that it naturally falls into three pieces
\begin{equation}\label{omega_gen}
    \Omega=\Omega_0+\Omega_q,\qquad\Omega_q=\Omega_{c}+\Omega_{os}.
\end{equation}
Here $\Omega_0$ is the quasiclassical (main) contribution. A discreteness of quantum numbers is completely neglected in this term. It is that contribution which is described by the naive heat-kernel expansion. The term $\Omega_q$ is an essentially quantum contribution. It is decomposed on the contribution from the branch cut of the dispersion law \eqref{disper law} (if exists) and the oscillating term $\Omega_{os}$ which is given by the sum over the Matsubara frequencies. The latter term oscillates with the chemical potential, while the former does not. Notice that, in this paper, we do not regard the problem of a vacuum energy and the general expression \eqref{omega_gen} for the $\Omega$-potential is assumed not to include the divergent vacuum term. Nevertheless, as we shall see, the vacuum contribution can be found from the high-temperature expansion of \eqref{omega_gen}, and the Casimir term \cite{CasPold} is contained in the contribution $\Omega_{c}$.

In Sec. \ref{One-loop_Omega}, we develop a general procedure for obtaining of rapidly convergent expansions of the partition function. We start with the definition of the $\Omega$-potential for particles (boson or fermions) with the dispersion law \eqref{disper law} and, by use of the Poisson summation formula, bring it to the form \eqref{omega_gen}. Then, every term in the representation \eqref{omega_gen} is studied separately. The oscillating term $\Omega_{os}$ reduces to a sum of the Epstein $\zeta$-functions \cite{Epstein} and, in Sec. \ref{Exp_Osc_Ter}, we, in fact, rederive the Chowla-Selberg expansion \cite{ChowSel,HJKS,Elizalde,LimTeo,AmbWolf,ElOdSah,Edery,Berndt,KTTY,KirstBo} for it. Contrary to the standard $\zeta$-function approach to evaluation of the one-loop corrections, we do not assume the CPT-symmetry (an identity of the particle and anti-particle dispersion laws) in our calculations. In this section, we also single out the contribution from the cut and establish its relation to the Casimir term and the vacuum energy. In Sec. \ref{Exp_Quas_Pol}, we represent the quasiclassical contribution $\Omega_0$ in the form analogous to that for the essentially quantum contribution $\Omega_q$, but assuming an analytical continuation in the parameter $d$, which initially characterizes the space dimension. In the next Sec. \ref{Exp_Eff_Cond}, we analyze an effectiveness of the obtained expansion and derive the restrictions on the spectrum parameters under which the expansion converges rapidly. These conditions also allow us to find a region in the parameter space where the oscillations become significant. In order to extend a range of applicability of the expansion, we further study its various resummations. Sec. \ref{Resum} is devoted to this problem. Here we consider three cases: i) the poles corresponding to the Matsubara frequencies merge and constitute a cut; ii) the problem reduces effectively to a one-dimensional one like, for example, in the case of a thin film with macroscopic transversal dimensions; iii) one or several poles approach the real axis so that their contributions to the partition function are not exponentially suppressed. In Sec. \ref{Quasi_Contr}, we turn back to the quasiclassical contribution $\Omega_0$ and derive its low- and high-temperature expansions. In particular, the obtained high-temperature expansion generalizes the known one  \cite{KirstHKE,polylog} for the relativistic dispersion law to the dispersion law of the form \eqref{disper law}. Here we also explicitly write out the finite and logarithmically divergent contributions to the one-loop vacuum energy for particles with the dispersion law \eqref{disper law}. It turns out that the logarithmic divergencies which are important for the renormalization group method arise only for certain dispersion laws, the relativistic spectrum being one of them.

In Sec. \ref{Examp}, we apply the general formalism to two models: massless relativistic particles (a linear dispersion law) in a box Sec. \ref{Massl_Part}, and  non-relativistic particles in a thin metal film Sec. \ref{Ele_Th_Me_Fi}. The latter model is a classical subject of the Fermi liquid theory of conduction electrons in metals and was studied in many papers and books (some of them are \cite{BregZhuch,Rumer,Zilb,LifshKos,Shoenb,GJPVW,Nedore,Azbel,Kulik,LifshKag}). As it was mentioned, our purpose in this case is to obtain a rapidly convergent expansion more accurate than the quasiclassical result of the general theory of oscillations \cite{LifshKag}. Here we obtain the closed simple expressions for the $\Omega$-potential and the average number of conduction electrons which are valid with an exponential accuracy. Also in Sec. \ref{Ele_Th_Me_Fi}, we consider a dependence of the chemical potential on the thickness of the metal film. It is known \cite{Nedore,Kulik,LifshKag} that the partition function oscillates when one changes the sizes of the system. We derive the simple expressions for a period and amplitude of these oscillations. The first model is related, for example, to photons confined to the ideal metal rectangular parallelepiped \cite{CasPold,AmbWolf} when the chemical potential vanishes, and to electrons near the Dirac points \cite{Wallace,McCl} in graphene (graphite) when the chemical potential is not zero. Despite a study of these subjects also has a long history, they remain hot nowadays. The investigation of thermodynamic properties of massless particles in pistons is provoked by the experimental observations of the Casimir force (see for review  \cite{Mostep}), while the second subject is concerned with graphene \cite{NGMJKGDF,KatNov,NGPNG,FuPiGoMo}.

In considering the conduction electrons in graphene and a thin metal film, we make certain approximations. First, we completely ignore an interaction of the electrons with the electromagnetic field not produced by the core electrons and nuclei, the latter being taken into account only in the form of the dispersion law.  This is a standard approximation. A theory which includes this interaction both at zero (see, e.g., \cite{VshiKli,CarmUll}) and finite temperature \cite{SharGus} is known. In connection with the Casimir effect, its path-integral formulation was recently given in \cite{BFGV} in the Dirac approximation (relativistic dispersion law). Second, describing the oscillations of the chemical potential due to a variation of sizes of a graphene specimen, we suppose that the additional contributions to it resulting from a deformation of the lattice \cite{SoCoLo,GuLiZho,NGPNG} are approximately constant. This changes are rather small for the crystal sizes and deformations we study. Moreover, we can simply add them to the electronic chemical potential relaxing thereby the constancy assumption. Some experimental investigations of the graphene properties under pressure where the deformations of its lattice are realized can be found in \cite{PGNLCH}. We make the same assumption in studying the oscillations in a thin metal film. Here this approximation is a standard one since a variation of the dispersion law of conduction electrons is small under reasonable deformations of the specimen. Third, we do not include in our consideration any disorder effects supposing that the considered crystals are close to ideal. In a certain approximation, these effects are taken into account by a convolution of the obtained expressions for the $\Omega$-potential over the chemical potential with the Lorentz distribution \cite{SharGus}.

It will be convenient to use the following notation. We denote by $\mu$ and call it the dimensionless chemical potential the quantity $\beta\tilde{\mu}$, where $\tilde{\mu}$ is a customary chemical potential. Besides, the $\Omega$-potential is defined as
\begin{equation}
    \beta\Omega(\beta,\mu,\xi):=\ln Z(\beta,\mu,\xi),\qquad Z:=\Sp e^{-\beta\hat{H}+\mu\hat{N}},
\end{equation}
where $\hat{H}$ is the many-particle Hamiltonian, $\hat{N}$ is a conserved charge and $\xi$ some other parameters characterizing the spectrum of the Hamiltonian.

\section{One-loop $\Omega$-potential}\label{One-loop_Omega}

In this section, we derive various rapidly convergent expansions of the series
\begin{eqnarray}\label{part func}
    I_d=\sum\limits_{p}\ln(1+e^{\mu-E(p)}), \quad p\in\Z^d,
\end{eqnarray}
with the dispersion law \eqref{disper law}. The sum in Eq. \eqref{part func} is taken over all the integer $p_i$, and, for convenience, we define the dimensionless chemical potential $\mu$ in such a way that $E(p)\geq0$ for any $p$. In the case we study, the one-loop contribution to the $\Omega$-potential is proportional to this series or expressed in terms of the series of such a type. Usually, the series \eqref{part func} corresponds to the partition function of quantum fields subjected to the periodic boundary conditions. As for bosons, the logarithm of the partition function differs from Eq. \eqref{part func} by the overall sign and by the sign at the exponent in the logarithm. The latter sign can be simply earned by the replacement $\mu\rightarrow\mu-i\pi$. Making use of the representation of the $\delta$-function in terms of the Fourier series, we can write (the Poisson formula)
\begin{equation}\label{Id_gen}
    I_d=\sum_q\int
    d^dp e^{-2\pi ip_iq^i}\ln(1+e^{\mu-E(p)})=\sum_qe^{2\pi ia_iq^i}\int d^dp e^{-2\pi ip_iq^i}\ln(1+e^{\mu-\omega(p^2+m^2)}),
\end{equation}
where $q$ is a multiindex taking the same values as the multiindex $p$.

\subsection{Expansion of the oscillating term}\label{Exp_Osc_Ter}

Let us consider in detail the integral
\begin{equation}\label{f_y}
    I_d^q=\int d^dp e^{-2\pi ip_iq^i}\ln(1+e^{\mu-\omega(p^2+m^2)})=:\int d^dp e^{-2\pi ip_iq^i}\tilde{f}(p^2).
\end{equation}
At first, we reduce it to the one-dimensional integral
\begin{equation}\label{onedim_int}
    I_d^q=2\pi\sqrt{g}\int_0^\infty dppj_{d/2-1}(q,p)\tilde{f}(p^2),
\end{equation}
where $g:=\det{g_{ij}}$, $p:=\sqrt{p^2}$ and $q:=\sqrt{q^2}$. Hereinafter, we use a convenient notation for the expressions containing the Bessel function $Z_\nu(x)$
\begin{equation}\label{Bessel_func}
    z_\nu(q,p):=\left(\frac{p}{q}\right)^\nu Z_\nu(2\pi qp).
\end{equation}
In particular, these functions obey the useful recurrence relations
\begin{equation}\label{recurr relat}
    \frac{\partial z_\nu(q,p)}{p\partial p}=2\pi z_{\nu-1}(q,p),\qquad \frac{\partial z_\nu(q,p)}{q\partial q}=-2\pi z_{\nu+1}(q,p).
\end{equation}
The integral $I_d^q$ is an analytic function of the complex variable $d$ at $\re d>0$. Besides, we shall assume that
\begin{equation}\label{bound_cond_mu}
    \lim_{\mu\rightarrow-\infty}I^q_d=\lim_{\mu\rightarrow-\infty}\partial_\mu I^q_d=0.
\end{equation}
This conditions will allow us to restore the function $I_d^q$ by its second derivative. The analytic continuation of the function $I_d^q$ to the whole complex $d$-plane can be achieved by the standard trick. We pass from the integral over the real semi-axis to the integral over the Hankel contour
\begin{equation}\label{onedim_int_anal}
    I_d^q=\frac{\pi\sqrt{g}}{e^{i\pi d}-1}\int_Hdsj_{d/2-1}(q,s^{1/2})\tilde{f}(s),
\end{equation}
where the contour $H$ runs from $+\infty$ a little bit higher the real axis, encircles the origin and then goes to $+\infty$ a little bit lower the real axis. When $d$ is an even number, we have to take the limit in Eq. \eqref{onedim_int_anal}. We also assume that we can choose such the branch of the function $\tilde{f}(s)$ which has not a cut on the real semi-axis. For any reasonable $\omega(s)$ the integral \eqref{onedim_int_anal} converges uniformly with respect to $\mu\leq\mu_0<0$. Then we see from the representation \eqref{onedim_int_anal} that the ``boundary'' conditions \eqref{bound_cond_mu} are satisfied for any $d\in\mathbb{C}$.

Consider the function
\begin{equation}\label{f_s}
    f(p^2):=\partial_\mu^2\tilde{f}(p^2)
\end{equation}
as an analytic function of the variable $s:=p^2$ in the complex plane. We shall suppose that the function $\omega(s+m^2)$ defining the dispersion law is analytic on the whole $s$-plane or possesses the branch point $s_c\leq 0$ on the real axis. A generalization to the case when the function $\omega(s+m^2)$ owns more than one branch points will be obvious. It is useful to choose such a branch of $\omega(s+m^2)$ which respects the Schwarz's symmetry principle
\begin{equation}\label{Schwarz_sym}
    \omega^*(s)=\omega(s^*).
\end{equation}
Then the function $f(s)$ also has this property. The cut picking out such a branch of $\omega(s)$ goes along the real axis from $-\infty$ to $s_c$. In addition to the mentioned branch point, the function $f(s)$ has singularities (the poles of the second order) in the points
\begin{equation}\label{zeros}
    \omega(s_k+m^2)=\mu+i\omega_k,\qquad \omega_k=\pi(2k+1),\quad k\in\Z,
\end{equation}
where $\omega(s)$ denotes the branch chosen by us and $\omega_k$ are the Matsubara frequencies. The solutions $s_k$ of Eq. \eqref{zeros} can coincide at different $k$. Moreover, Eq. \eqref{zeros} can have a non-unique solution at the fixed $k$ or have not any one. Due to the symmetry \eqref{Schwarz_sym}, the poles $s_k$ appear by the complex conjugate pairs. Further, we shall denote by $s_k$ all the different solutions to Eq. \eqref{zeros}.

Let $C'$ be the contour enclosing the poles \eqref{zeros} in a counterclockwise direction and $C_\infty$ be the circle with a radius tending to infinity. Then, by the use of the Cauchy formula, $f(p^2)$ can be represented in a form analogous to the K\"{a}llen-Lehmann representation for the Fourier transform of an exact propagator
\begin{equation}\label{kall-lehm}
    f(p^2)=-\int\limits_C \frac{ds}{2\pi i}\frac{f(s)}{p^2-s}=\int\limits_{C'} \frac{ds}{2\pi i}\frac{f(s)}{p^2-s}-\int\limits_{-\infty}^{s_c}\frac{ds}{2\pi i}\frac{f(s+i\epsilon)-f(s-i\epsilon)}{p^2-s}-\int\limits_{C_\infty} \frac{ds}{2\pi i}\frac{f(s)}{p^2-s},
\end{equation}
where the contour $C$ encloses only the singular point $s=p^2$, and $\epsilon$ is an infinitesimally small positive quantity. Also we assume that $f(s)$ is integrable in a vicinity of the branch point $s=s_c$. If the function $\omega(s)$ has no branch points, the second integral should be omitted. If $f(s)$ tends to zero almost everywhere at $|s|\rightarrow\infty$ then the last integral is also absent. In particular, the last property is fulfilled for the function \eqref{f_s} if
\begin{equation}
    \lim_{|s|\rightarrow\infty}|\omega(s)|=\infty.
\end{equation}
Otherwise, we need to make subtractions in Eq. \eqref{kall-lehm}. Discarding the integral over the large circle in \eqref{kall-lehm}, we arrive at
\begin{multline}\label{integral I}
    \partial_\mu^2I_d^q=2\pi\sqrt{g}\int_0^\infty dppj_{d/2-1}(q,p)\left[\int\limits_{C'} \frac{ds}{2\pi i}\frac{f(s)}{p^2-s}-\int\limits_{-\infty}^{s_c} \frac{ds}{\pi}\frac{\im f(s+i\epsilon)}{p^2-s}\right]\\
    =-i\sqrt{g}\left[\int\limits_{C'}ds f(s)-2i\int\limits_{-\infty}^{s_c} ds\im f(s+i\epsilon)\right]\int_0^\infty dpp \frac{j_{d/2-1}(q,p)}{p^2-s},
\end{multline}
The change of an integration order in the last equality is valid when $\re d\in(0,3)$ at $q\neq0$, and $\re d\in(0,2)$ at $q=0$. If the physical dimension does not fall into these intervals then we shall make all the calculations as it would be in these intervals and set its value to the real one in the final result only. These operations are legitimate by virtue of a uniqueness of an analytic continuation.

Further, we shall consider the case $q\neq0$. Notice that the integral factorized in \eqref{integral I} can be interpreted as the inverse Fourier transform of the free propagator with the mass $s^{1/2}$ in a $d$-dimensional Euclidean space. Now we apply the formula (\cite{Wats}, p. 434)
\begin{equation}\label{Hankel trans}
    \int_0^\infty dp p\frac{j_\nu(q,p)}{(p^2-s)^{\al+1}}=\frac{i\pi^{\al+1}}{2\Ga(\al+1)}h^{(1)}_{\nu-\al}(q,s^{1/2}),\qquad q>0, \quad\re\nu\in\left(-1,2\al+\frac32\right),
\end{equation}
and $s$ does not lie on the real positive semi-axis. The cut of the function $s^{1/2}$ is also along the real positive semi-axis so $s^{1/2}$ takes the values in the upper half-plane. At $\al=0$ we obtain
\begin{equation}\label{idqmu}
    \partial_\mu^2I_d^q=\frac{\sqrt{g}}2\biggl[\int\limits_{C'}ds f(s)\partial_sh^{(1)}_{d/2}(q,s^{1/2})-4\int\limits_{|s_c|}^\infty ds \im f(-s+i\epsilon)k_{d/2-1}(q,s^{1/2})\biggr],
\end{equation}
where we have used the recurrence relation \eqref{recurr relat} and the relation of the Hankel function with the Macdonald one
\begin{equation}\label{Hankel_MacDon}
    i\pi h^{(1)}_\nu(q,ip)=2k_\nu(q,p).
\end{equation}
It also follows from Eq. \eqref{Hankel trans} that
\begin{equation}\label{Hankel_conj}
    \left[h^{(1)}_\nu(q,s^{1/2})\right]^*=-h^{(1)}_\nu(q,(s^*)^{1/2}),
\end{equation}
when $\nu$ and $q$ are real.

Now we take into account an explicit form of the function $f(s)$. Then, for any function $\vf(s)$ analytic in a neighbourhood of the point $s_k$, the following relation holds
\begin{equation}\label{residue}
    \underset{s=s_k}{\res}[f(s)\vf(s)]=-\partial_\mu[\partial_\mu s_k\vf(s_k)],
\end{equation}
where $s_k$ are specified by Eq. \eqref{zeros}. Bearing in mind the boundary conditions \eqref{bound_cond_mu}, the integral $I_d^q$ is rewritten as the sum of contributions from the residues and the cut
\begin{multline}\label{expansion}
    I_d^q=\sqrt{g}\biggl[\int\limits_{|s_c|}^\infty dsk_{d/2-1}(q,s^{1/2})\im\left(\omega^+(s)-\ln\cosh^2\frac{\omega^+(s)-\mu}{2}\right)-i\pi\sum_kh^{(1)}_{d/2}(q,p_k)\biggr]\\
    =\sqrt{g}\im\biggl\{\int\limits_{|s_c|}^\infty ds\left[\omega^+(s)k_{d/2-1}(q,s^{1/2})+\frac{\omega'^+(s)}{\pi}\tanh\frac{\omega^+(s)-\mu}2k_{d/2}(q,s^{1/2})\right]
    +\pi\sum_kh^{(1)}_{d/2}(q,p_k)\biggr\},
\end{multline}
where we define $p_k:=s^{1/2}_k$ and the function $\omega^+(s):=\omega(m^2+i\epsilon-s)$. The summation is taken over all the residues \eqref{zeros} of the function $f(s)$. According to our definition of the square root, $\im p_k\geq0$ and, consequently, the contributions from the poles and the cut to the $\Omega$-potential are exponentially suppressed. In the last expression for $I_d^q$, the variable $d$ is supposed to be real. In this expression, we explicitly singled out the Casimir contribution from the cut to the logarithm of the partition function (the first term in the square brackets) and also integrated the reminder by parts assuming that the discontinuity of the integrand goes to zero at the branch point. A bosonic variant of formula \eqref{expansion} is produced by the substitution $\mu\rightarrow\mu-i\pi$ and change of the overall sign.

Some comments on the expansion \eqref{expansion} are in order. If the function $\omega(s)$ takes the values on the opposite banks of the cut which differ by a sign only, as, for example, for the relativistic dispersion law, then an imaginary part of the second term in the square brackets in the expansion \eqref{expansion} becomes an odd function of $\mu$. Therefore, if the considered model allows for the antiparticles and we include their contribution to the $\Omega$-potential, these terms cancel out and the contribution from the cut is given solely by the Casimir term. For certain dispersion laws and values of the chemical potential, the situation may occur when some poles in the $s$-plane appear precisely on the cut of the function $\omega(s+m^2)$. In this case, the integrals in \eqref{expansion} are understood in the sense of principal value, while the contribution from these poles enters the expansion \eqref{expansion} with the factor $1/2$.

Notice that, in order to obtain the contribution from the poles to the sum \eqref{Id_gen}, we could use the well-known representation of the Epstein $\zeta$-function in terms of a series in the Macdonald functions \cite{Elizalde,ChowSel,HJKS,LimTeo,AmbWolf,ElOdSah,Edery,Berndt,KTTY,KirstBo}. With this observation, we would come to the expansion \eqref{expansion} without the contribution from the cut at one stroke. Also, if the function $\omega(s)$ has no branch points, we could use the general formula \cite{Sahar} for integrals of the type \eqref{onedim_int} and obtain the expansion \eqref{expansion}.

Let us estimate a behaviour of the different contributions to \eqref{expansion} at high temperatures (small $\beta$). In the fermionic case, the contributions from the poles tends exponentially to zero for the dispersion laws $\omega(s)$ with the power-like asymptotics at high energies. This holds for bosons as well excepting the contribution from the pole with zero Matsubara frequency. At the fixed chemical potential $\tilde{\mu}\equiv\mu/\beta$, its contribution to the average energy is proportional to $\beta^{-1}$. In the fermionic case, the contribution of the second term in the square brackets tends linearly to zero at $\beta\rightarrow0$ and fixed $\tilde{\mu}$. For bosons, the contribution of this term to the average energy behaves like $\beta^{-1}$ at the fixed chemical potential $\tilde{\mu}$. The first term in the square brackets (the Casimir term) depends linearly on the temperature. So, its contribution to the average energy is independent of $\beta$. It is the Casimir term which determines the leading asymptotics in $\beta$ of the essentially quantum (i.e. non-quasiclassical) thermal correction to the average energy in the fermionic case. As far as the bosons are concerned, this leading asymptotics is determined by the above mentioned terms proportional to $\beta^{-1}$. At low temperatures and $E(p)>0$, all these terms are absorbed by other contributions to the average energy so that, in sum, it tends exponentially to zero.

The Casimir term is exactly canceled by the analogous term coming from the vacuum fluctuations. It is easy to understand from the following general observation. An expression for the one-loop vacuum energy can be obtained from the high-temperature expansion for fermions at the vanishing chemical potential since
\begin{equation}\label{zero_point_energ}
    E_f(\beta,0)=-\left.\partial_\beta(\beta\Omega_f(\beta,\mu))\right|_{\mu=0}=\sum_n\frac{E_n}{e^{\beta E_n}+1}\underset{\beta\rightarrow0}{\rightarrow}\sum_n\frac{E_n}{2},
\end{equation}
where $E_n$ is the energy of the mode $n$, the index $f$ marking the statistics. In this method for obtaining the vacuum energy, the Fermi-Dirac distribution plays the role of a regulator resulting in the energy cutoff. Let the contribution to the average energy at zeroth power of $\beta$ takes the form
\begin{equation}\label{zero_point_energ_expl}
    E_f(\beta,\mu)=a^0_0(\mu)+a^0_1(\mu)\ln(\beta b^0_1(\mu))+\ldots,
\end{equation}
where dots denote the terms at other powers of $\ln\beta$ and $\beta$. The quantities $a^k_l(\mu)$ and $b^k_l(\mu)$ depend on the spectrum parameters. The vacuum energy divergencies, appearing when $\beta$ tends to zero, should be canceled by appropriate counterterms in the initial action of the model. This renormalization procedure reduces, in essence, to a replacement of the divergent in the limit $\beta\rightarrow0$ coefficients at the different structures $a^k_l(0)\ln^nb_l^k(0)$, $n<l$ and $k\leq0$, by some finite constants. These constants have to be fixed by certain additional normalization conditions. Keeping in mind \eqref{zero_point_energ}, we see that the total renormalized average energy of the system of fermions with their vacuum at a finite temperature does not contain the term $a_0^0(0)$ in the following sense
\begin{multline}
    E^{tot}_f(\beta,\mu)=-\left.\sum_n\frac{E_n}{2}\right|_{\text{ren.}}+E_f(\beta,\mu)\\
    =a^0_0(\mu)-a^0_0(0)+(a^0_1(\mu)-a^0_1(0))\ln(\beta b^0_1(\mu))+a^0_1(0)\ln\frac{\beta b^0_1(\mu)}{cb^0_1(0)}+\ldots,
\end{multline}
where $c$ is some constant. This property holds for bosons as well, inasmuch as the vacuum energy can be obtained from the high-temperature limit of $E_b(\beta,0)-2E_b(2\beta,0)=E_f(\beta,0)$. However, it does not mean that the quantity $a^0_0(0)$ cannot be observed at the vanishing chemical potential. At low temperatures, the thermal contribution to the average energy tends exponentially to zero, whereas the vacuum energy is independent of the temperature and has the form \eqref{zero_point_energ_expl} with $\mu=0$ and renormalized (not necessarily to zero) divergencies.

To a ceratin extent, the above reasonings can be generalized to the multi-loop corrections with a regulator of the thermal form, but we leave a detailed study of this question for a future research.

\subsubsection{Expansion of the quasiclassical contribution in terms of poles}\label{Exp_Quas_Pol}

The quasiclassical contribution to the sum \eqref{Id_gen} at $q=0$ can be also represented in the form analogous to \eqref{expansion}. To this end, we put $q=0$ in the integral \eqref{integral I}. Then we obtain
\begin{equation}
    \partial_\mu^2I_d^0=-i\sqrt{g}\frac{\pi^{d/2-1}}{\Gamma(d/2)}\left[\int\limits_{C'}ds f(s)-2i\int\limits_{-\infty}^{s_c} ds\im f(s+i\epsilon)\right]\int_0^\infty dp\frac{p^{d-1}}{p^2-s},\quad\re d\in(0,2).
\end{equation}
The integral over $p$ reduces to the beta function. Taking the integral over $s$ and using the relation \eqref{residue}, we arrive at
\begin{equation}\label{contrib_q0_pre}
    \partial_\mu^2I_d^0=-\pi^{d/2}\sqrt{g}\Gamma(-d/2)e^{-i\pi d/2}\sum_k\partial_\mu^2p_k^d-\pi^{d/2-1}\sqrt{g}\Ga(1-d/2)\int\limits_{|s_c|}^{\infty}dss^{d/2-1}\im f(-s+i\epsilon),
\end{equation}
where the summation is carried over all the poles of the function $f(s)$. Note that this expression is real when the variable $d$ is real. It is, of course, an anticipated result. In order to take advantage of this formula one needs to sum the series over $k$, represent the result as an analytic function of $d$, and only then set $d$ to the physical dimension. The same procedure should be applied to the integral over the cut. Also, if it is necessary, one can deform the integration contour to make the integral convergent.

Integrating Eq. \eqref{contrib_q0_pre} over $\mu$, at that $d$ when the resulting series in the right-hand side (RHS) converges, and taking into account the boundary conditions \eqref{bound_cond_mu}, we can write
\begin{multline}\label{contrib_q0}
    I_d^0=-\pi^{d/2}\sqrt{g}\Gamma(-d/2)e^{-i\pi d/2}\biggl[\sum_kp_k^d(\mu)-\sum_kp_k^d(-\infty)\biggr]\\
    +\frac12\pi^{d/2-1}\sqrt{g}\Gamma(1-d/2)\int\limits_{|s_c|}^\infty ds\left[s^{d/2-1}\im\omega^+(s)+d^{-1}s^{d/2}\im\left(\omega'^+(s)\tanh\frac{\omega^+(s)-\mu}{2}\right)\right],
\end{multline}
where we have integrated by parts like in Eq. \eqref{expansion}. If $|p_k(\mu)|$ tends to infinity with $\mu$ tending to minus infinity then the last term in the square brackets in the first line vanishes. This is the case, for example, when the dispersion law $\omega(s)$ has a power-like asymptotics at $|s|\rightarrow\infty$. Just as for a contribution from the cut to the essentially quantum part of the $\Omega$-potential, $\Omega_q$, the second term in the square brackets disappears when the values of the function $\omega(s)$ taken on the opposite banks of the cut differ by a sign only and we take into account the contribution from antiparticles.

\subsection{Expansion effectiveness conditions}\label{Exp_Eff_Cond}

In order to clarify the conditions of an effectiveness of the expansion \eqref{expansion}, it is useful to provide it by a simple visual interpretation. As we have already mentioned, every term of the series over poles in Eq. \eqref{expansion} substituted to \eqref{Id_gen} can be thought of as the Euclidean propagator of a particle with the mass $\re p_k$ and the rate of decay $\im p_k$ in $(d+2)$-dimensional space. The sum entering the expansion \eqref{expansion} can be interpreted as a sum of wave-functions (taken in the origin) of these ``particles'' propagating from the lattice points with coordinates $q^i$ and the initial phases $e^{2\pi i a_iq^i}$. This lattice is embedded in a $(d+2)$-dimensional Euclidean space, whereas the Gram matrix of the lattice spacings specifies the metric $g_{ij}$.

Now it is evident that the sum over poles converges rapidly if the lattice spacings are large enough, i.e., the elements of the Gram matrix $g_{ij}$ are large, and the ``screening parameter'' $\im p_k$ tends rapidly to infinity with the number $k$. Then, with a neglible error, we can break off the series in $k$ and $q$ retaining only several first terms in them. We see from Eq. \eqref{zeros} defining the singular points that the imaginary part of $p_k$ grows with $k$ for $\omega(s)\sim s^\al$, $\al>0$, at large $s$, it growing slower at the greater $\al$ (with the exception of $\al=1/4n$). If $\omega(s)\sim\exp(\al s)$, the imaginary part of $p_k$ decreases. Thus, if the dispersion law $\omega(s)$ grows with $s$ more rapidly than any power of $s$ then the expansion \eqref{expansion} cannot be used to make an accurate estimation of the sum \eqref{part func}\footnote{If the function $\omega(s)$ grows rapidly with $s$, a good approximation for the partition function can be obtained directly from its definition \eqref{part func} discarding all the higher terms of the sum. Analogously, we can obtain a good approximation for the partition function in the case of small metric elements $g_{ij}$.}. Further, we shall assume that $\im p_k$ and $|p_k|$ increase monotonically when an absolute value of the number $k$ increases.

To derive a rough estimate of the expansion effectiveness conditions, we require that a term of the series in $k$ with the number $k_0$ will be much greater than the next term of this series starting with some small number $k_0$. Let us use the series representation of the Hankel function at a large argument \cite{GrRy}
\begin{equation}\label{Hankel_func_expans}
    h^{(1)}_\nu(q,p)\approx\pi^{-1} e^{i\pi(2qp-\nu/2-1/4)}\sum_{n=0}^\infty\left(\frac{i}{4\pi}\right)^n\frac{p^{\nu-n-1/2}}{q^{\nu+n+1/2}}\frac{\Ga(\nu+n+1/2)}{n!\Ga(\nu-n+1/2)}.
\end{equation}
This is an asymptotic expansion. If we take only the first $M$ terms of this series, we make an error with an absolute value lesser than an absolute value of the first discarded term. The reminder term has the same form as a general term of the series taken at $n=M+1$ and multiplied by a number with a modulus lesser than one. For a half-integer $\nu$, this series terminates giving rise to a closed expression for the Hankel function in terms of elementary functions. Using this asymptotic expansion, we have from \eqref{expansion} the condition of a rapid convergence
\begin{equation}\label{exp_cond}
    2\pi q_m\im(p_{k_0+1}-p_{k_0})-\frac{d}4\ln\left|\frac{p^2_{k_0+1}}{p^2_{k_0}}\right|\gtrsim1,
\end{equation}
where $q_m:=\min(q)$, $q\neq0$. This condition constrains the values of the metric components $g_{ij}$ which characterizes the sizes of a system. In particular, this condition implies  $2\pi q_m\im p_{k_0+1}>1$, what allows us to break off the series in $q$ for the contributions from the poles with numbers $k_0+1$ and higher. In order to terminate the sum over $q$ for the contributions from the poles $p_k$, $|k|\leq k_0$, it is necessary to demand, in addition,
\begin{equation}\label{exp_cond_q}
    2\pi q_m\im p_k\gtrsim1,\qquad\text{for } |k|\leq k_0.
\end{equation}
We shall show below how to obtain an expansion of the logarithm of the partition function in that case when the last condition violates for a few number of poles. Notice also that the expansion effectiveness conditions can be used to specify a domain of values of the system parameters in which the oscillating non-extensive corrections described by the pole terms appreciably contribute to the $\Omega$-potential. This domain is at a saturation border of the inequality \eqref{exp_cond_q}.

As one can see from \eqref{Hankel_func_expans} and the interrelation between the Macdonald and Hankel functions \eqref{Hankel_MacDon}, the sum over $q$ in the contribution from the cut to the expansion \eqref{expansion} can be broken off if
\begin{equation}
    2\pi q_m |s_c|^{1/2}\gtrsim1.
\end{equation}
The method of obtaining of a rapidly convergent expansion in the case when the branch point $s_c$ approaches zero will be considered below.

The expansion effectiveness conditions can be written more explicitly if we know a concrete form of the dispersion law. For example, in the relativistic case, the poles defining the expansion of the $\Omega$-potential becomes
\begin{equation}\label{roots rel}
    p_k=\sqrt{(\mu+i\omega_k)^2-m^2},\qquad\im p_k>0,\quad k\in\Z.
\end{equation}
Recalling our choice for the branch of a square root entering the dispersion law, we see that $\mu$ should be positive. When we decrease the value of $\mu$, the poles in the $s$-plane approach the cut $(-\infty,-m^2]$ from the above and below in a symmetric way. At $\mu=0$ all the poles lie on the cut and have to be taken with the factor $1/2$ in the expansion \eqref{expansion}. At $\mu<0$ all the poles leave the ``physical'' sheet and the sum over them in \eqref{expansion} vanishes. From the physical point of view, this corresponds to the fact that, at negative chemical potentials, particles and antiparticles switch their roles -- the probability for a particle to appear in the system becomes exponentially suppressed.

At sufficiently small temperatures ($m$ and $\mu$ are much greater than $\omega_{k_0}$), the condition \eqref{exp_cond} approximately reads as
\begin{equation}\label{cond rel}
\begin{aligned}
    4\pi^2 q_m \mu&\gtrsim(\mu^2-m^2)^{1/2}, &\quad \text{$\mu>m$;}\\
    2\pi^2 q_m(\omega_{k_0+1}+\omega_{k_0})m^2&\gtrsim(m^2-\mu^2)^{3/2}, &\quad \text{$\mu<m$.}
\end{aligned}
\end{equation}
As for bosons, we need to use the second condition. In the Boltzmann limit, $\mu<m$ for fermions as well (see, e.g., \cite{LandLifshStat}), although, for a degenerate Fermi gas, the nonrelativistic chemical potential $\mu_{nr}=\mu-m$ is positive. At high temperatures ($m$ and $\mu$ are small in comparison with $\omega_{k_0}$), the condition \eqref{exp_cond} in the leading order takes the form
\begin{equation}\label{cond rel_high}
    4\pi^2q_m\gtrsim1+\frac{d}2\ln\left|\frac{\omega_{k_0+1}}{\omega_{k_0}}\right|.
\end{equation}
This condition is, in fact, the expansion effectiveness condition \eqref{expansion} for the oscillating part of the partition function in the case of a massless quantum field at zero chemical potential.

For the nonrelativistic dispersion law, the poles look like
\begin{equation}
    p_k=\sqrt{\mu+i\omega_k},\qquad \im p_k>0,\quad k\in\Z.
\end{equation}
At low temperatures ($\mu$ is much greater than $\omega_{k_0}$), the condition \eqref{exp_cond} in the leading order reduces to
\begin{equation}
\begin{aligned}
    2\pi^2q_m&\gtrsim\sqrt{\mu},&\quad\text{$\mu\gg1$;}\\
    \pi^2 q_m(\omega_{k_0+1}+\omega_{k_0}) &\gtrsim2|\mu|^{3/2},&\quad\text{$\mu\ll-1$.}
\end{aligned}
\end{equation}
At high temperatures, the condition \eqref{exp_cond} becomes
\begin{equation}
    \pi q_m\gtrsim\frac{1+\ln|\omega_{k_0+1}-\omega_{k_0}|^{d/4}}{\sqrt{2\omega_{k_0+1}}-\sqrt{2\omega_{k_0}}}\left(1-\frac{\mu}{2\sqrt{\omega_{k_0+1}\omega_{k_0}}}\right),
\end{equation}
where we have taken into account the first correction in $\mu$. One can also derive these conditions (without the first correction) as a nonrelativistic limit of \eqref{cond rel} and \eqref{cond rel_high}.

\subsection{Resummations}\label{Resum}

\subsubsection{Summation over poles}

Let there exist a sequence $s_\al$, $\al=\overline{N,\infty}$, among the poles $s_k$ such that the distance between adjacent members of this sequence $|p_{\al+1}-p_\al|$ tends to zero with an increase of the number $\al$, whereas $\im p_\al$ tends to infinity. Then, the estimate of the $\Omega$-potential, which is obtained by taking into account only the first several terms of the expansion \eqref{expansion}, can be improved summing the sequence $s_\al$ by the Euler-Maclaurin formula
\begin{equation}
    \sum_{\al=N}^\infty h^{(1)}_{d/2}(q,p_\al)=\int_{s_N}^\infty\frac{ds}{2\pi^2i}\omega'(s+m^2)\partial_sh^{(1)}_{d/2+1}(q,s^{1/2})+\frac12h^{(1)}_{d/2}(q,p_N)+\ldots,
\end{equation}
where the integration is along the contour which connects the poles $s_\al$ starting from the $N$-th pole. In consequence of the symmetry \eqref{Schwarz_sym}, the sequence of conjugate poles $s^*_\al$ also contributes to the expansion \eqref{expansion}. Keeping only the leading integral term, we obtain for a contribution of these two sequences
\begin{equation}\label{Idq_Eu_Mac}
    -\sqrt{g}\re\int_{s_N}^\infty \frac{ds}{\pi}\omega'(s+m^2)\partial_sh^{(1)}_{d/2+1}(q,s^{1/2})+\ldots
\end{equation}
The integral in the Euler-Maclaurin formula can be taken for a nonrelativistic dispersion law when $\omega'(s)=1$ and the distance between the poles $p_k$ tends to zero at $k\rightarrow\infty$. This holds for any polynomial spectrum as well. Formula \eqref{Idq_Eu_Mac} resembles the contribution from the cut entering the expansion \eqref{expansion}. It is not surprising since the poles merge and constitute a cut when a distance between them is small.

\subsubsection{Summation over $q$ in one dimension}

At the large argument,
\begin{equation}\label{Hankel_func_arg}
    4\pi q|p|\gg1,
\end{equation}
the Hankel function admits the asymptotic expansion \eqref{Hankel_func_expans}. If the summation over $q^i$ in the series \eqref{Id_gen} reduces effectively to a one-dimensional sum, viz.
\begin{equation}
    q=g^{1/2}_{11} |q^1|=q_m|q^1|,
\end{equation}
then we can sum the series over $q$
\begin{multline}\label{onedim_sum}
    \sideset{}{'}\sum_{q_1}e^{2\pi i a_1q^1}h^{(1)}_\nu(q,p)\approx\pi^{-1}e^{-i\pi(\nu/2+1/4)}\sum_{n=0}^\infty\left(\frac{i}{4\pi}\right)^n\frac{p^{\nu-n-1/2}}{q_m^{\nu+n+1/2}}\frac{\Ga(\nu+n+1/2)}{n!\Ga(\nu-n+1/2)}\\
    \times\left[\Li_{\nu+n+1/2}(e^{2\pi i(q_mp+a_1)})+\Li_{\nu+n+1/2}(e^{2\pi i(q_mp-a_1)})\right],
\end{multline}
where we have introduced the polylogarithm function
\begin{equation}\label{polylog def}
    \Li_\nu(z)=\frac{z}{\Gamma(\nu)}\int_0^\infty dx\frac{x^{\nu-1}}{e^x-z},\qquad\Li_\nu(z)=\sum_{k=1}^\infty\frac{z^k}{k^\nu},\quad|z|<1.
\end{equation}
The above described situation occurs, for example, in a one-dimensional problem or in the case of a thin film. In the latter case, the summation is carried over that quantum number which corresponds to the small dimension of a film. The contributions of transverse dimensions to the oscillating part of the $\Omega$-potential are strongly exponentially suppressed. The obtained expansion \eqref{onedim_sum} can be also used to evaluate the contribution to the partition function of the integral \eqref{Idq_Eu_Mac} coming from the Euler-Maclaurin formula or the integral representing the contribution from the cut. In order to apply it in these cases, the argument of the Hankel function should be large enough on the integration curve, that is, either the condition \eqref{Hankel_func_arg} is fulfilled for the pole $p_N$ at $q=q_m$ or the Hankel function has a half-integer index.

By the same way, we can sum over $q$ along any vector on the lattice in the $q$-space. This allows us to rewrite the whole sum over $q$ as the sum over all irreducible vectors, whose components have not a common divisor excepting unity, in the $q$-space. As a result, we obtain a sum of the expressions of the form \eqref{onedim_sum} over all the irreducible vectors of an increasing length. This resummation is an effective tool in the ``one-dimensional'' case or in the case when the condition \eqref{exp_cond_q} is weakly violated for the given pole $p_k$.

\subsubsection{Contribution of a pole approaching the real axis}\label{Contr_Pole_Real_Axis}

Let us now study the case when the condition \eqref{exp_cond_q} is not satisfied by some pole and the series in $q$ cannot be terminated. This case takes place when that pole approaches the real positive semi-axis in the $s$-plane
\begin{equation}\label{pole}
    s_0=b+i\epsilon,\qquad b\geq0,
\end{equation}
and $\epsilon$ is a tiny quantity. Besides, we can assume that $q_m$ is not small since, in the opposite case, the partition function in its initial form \eqref{part func} rapidly converges and can be simply estimated by the first several terms of the sum. Usually, the poles approach the real axis at low temperatures (see Eq. \eqref{zeros}) or, for the Bose-Einstein distribution, at a small chemical potential like in the case of a condensation. In spite of the fact that, in this section, we are about to investigate the pole contribution, all the analysis presented below is also applicable to the Macdonald function entering the integral over the cut in \eqref{expansion}, when the branch point $s_c$ is in close proximity to the origin.

Consider separately the contribution from the pole \eqref{pole} to the oscillating part of the partition function  \eqref{Id_gen}
\begin{equation}\label{contrib_one_pole_osc}
    c_{os}:=-i\pi\sqrt{g}\sideset{}{'}\sum_qe^{2\pi i a_iq^i}h^{(1)}_{d/2}(q,p_0).
\end{equation}
Recalling the integral representation of the Hankel function \cite{GrRy},
\begin{equation}
    i\pi h^{(1)}_{d/2}(q,p_0)=\pi^{d/2}\int_0^\infty idt (it)^{-1-d/2}e^{i(\pi^2 q^2/t+s_0 t)},\qquad\im s_0>0,\;\re d<0,
\end{equation}
the recurrence relation \eqref{recurr relat} and the symmetry property \eqref{Hankel_conj}, we see that $h^{(1)}_\nu(q,p)$ is an entire function of the variable $\nu$ at $q\neq0$ and $p\not\in\mathbb{R}$. As long as the series \eqref{contrib_one_pole_osc} converges uniformly with respect to $d\in\mathbb{C}$, it defines an entire function of the variable $d$. It is convenient to consider this series at $\re d<0$ when we can add the term $q=0$ to the sum. In this case,
\begin{equation}\label{contrib_one_pole}
    c:=c_{os}-\pi^{d/2}\sqrt{g}(-s_0)^{d/2}\Ga(-d/2)=-i\pi\sqrt{g}\sum_qe^{2\pi i a_iq^i}h^{(1)}_{d/2}(q,p_0)=-\sum_p\pi^{-\mu}\Ga(\mu)(P^2-s_0)^{-\mu},
\end{equation}
where $P_i:=p_i+a_i$ and $\mu:=(\bar{d}-d)/2$, while $\bar{d}$ is the physical space dimension. Regarding $\mu$ as an arbitrary parameter, we shall derive the rapidly convergent expansion of the contribution $c(\mu)$ from the one pole. Then we shall set $\mu$ to zero. For a definiteness, we shall assume that $\epsilon>0$. The expansion for the negative $\epsilon$ can be obtained by a complex conjugation. So,
\begin{equation}\label{contrib_one_pole_c}
    c=-\pi^{d/2}\sqrt{g}\sum_q\int_0^\la idt (it)^{-1-d/2}e^{i(\pi^2 q^2/t+s_0 t+2\pi a_iq^i)}-\pi^{-\mu}\sum_p\int_\la^\infty idt(it)^{\mu-1}e^{-it(P^2-s_0)},
\end{equation}
where $\la$ is some complex number. It is useful to choose $\la$ to lie on the imaginary axis, $\la=-i|\la|$, although it is not obligatory. In the first integral, we pick out the term $q=0$,
\begin{equation}\label{contrib_q0_one_pole}
    c_1^0:=-\pi^{d/2}\sqrt{g}(-s_0)^{d/2}[\Ga(-d/2)-\Ga(-d/2,-|\la| s_0)],
\end{equation}
where $\Ga(\nu,z)$ is an incomplete gamma function \cite{GrRy}
\begin{equation}\label{Gamma_func_inco}
    \Ga(\nu,z)=\Ga(\nu)-\sum_{n=0}^\infty\frac{(-1)^nz^{\nu+n}}{n!(\nu+n)}.
\end{equation}
The first term in the expression for $c_1^0$ is exactly the contribution to the sum \eqref{contrib_one_pole} at $q=0$. This term does not enter $c_{os}$. The contribution from $q\neq0$ to the first integral in Eq. \eqref{contrib_one_pole_c} can be written as
\begin{multline}\label{contrib_q_one_pole}
    \tilde{c}_1:=-\pi^{d/2}\sqrt{g}\sideset{}{'}\sum_q e^{2\pi ia_iq^i}\sum_{n=0}^\infty\frac{(\pi^2q^2 s_0)^n}{n!(\pi q)^{d}}\Ga(d/2-n,\pi^2q^2/|\la|)\\
    \approx-\frac{\pi^{d/2}}{|\la|^{d/2}}\sqrt{g}\sideset{}{'}\sum_q e^{-\pi^2 q^2/|\la|+s_0|\la|+2\pi ia_iq^i}\sum_{n=0}^\infty\left[\left(\frac{-1}{S'(t)}\frac{d}{dt}\right)^n\frac{t^{-1-d/2}}{S'(t)}\right]_{t=1},
\end{multline}
where $S'(t):=\pi^2q^2|\la|^{-1}t^{-2}+s_0|\la|$. The expansion in the first line should be exploited when the quantity $\pi^2 q^2|s_0|$ is small, while the asymptotic WKB-expansion in the second line is useful when $|S'(1)|$ is large. The second integral in Eq. \eqref{contrib_one_pole_c} is expressed through the incomplete gamma function
\begin{equation}\label{contrib_p_one_pole}
    c_\infty:=-\pi^{-\mu}\sum_p(P^2-s_0)^{-\mu}\Ga\left(\mu,|\la|(P^2-s_0)\right).
\end{equation}

Let us remind some properties of the incomplete gamma function which are needed for our subsequent investigation. The incomplete gamma function $\Ga(\nu,z)$ is an entire function of $\nu$ at $z\neq0$. In the $z$-plane, we choose its branch with a cut along the negative real semi-axis for $\nu\not\in\mathbb{N}$. The incomplete gamma function $\Ga(\nu,z)$ is singular in the point $z=0$ at $\re\nu\leq0$. Its absolute value grows exponentially at $\re z\rightarrow-\infty$ and exponentially goes down to zero at $\re z\rightarrow\infty$. In particular, we see that the contribution from the one pole to the oscillating part of the sum \eqref{Id_gen},
\begin{equation}
    c_{os}=\tilde{c}_1+c_\infty+c^0_1+\pi^{d/2}\sqrt{g}(-s_0)^{d/2}\Ga(-d/2),
\end{equation}
is an entire function of $\nu$ as expected. Now, we can put $\mu=0$.

The parameter $\la$ is an arbitrary number and resembles the massive parameter $\mu$ of the dimensional regularization of quantum field theory. We choose the value of the parameter $\la$ so that the sums over $q$ and $p$ in the expressions for $\tilde{c}_1$ and $c_\infty$ can be broken off with a neglible error. In order to terminate the sum over $q$ in \eqref{contrib_q_one_pole}, it is necessary to demand
\begin{equation}\label{cond_la_1}
    \pi^2 q^2_m/|\la|\gtrsim1.
\end{equation}
Then the term $\tilde{c}_1$ is exponentially small. At the same time, $\la$ should not be very small since, in that case, the sum \eqref{contrib_p_one_pole} over $p$ converges slowly. The main contribution to the sum over $p$ comes from the points in a vicinity of the ``resonance'' $P^2=b$ and from the exponentially growing contributions at a large negative real part of the argument of the incomplete gamma function. By use of the expansion \eqref{Gamma_func_inco}, it is not difficult to estimate a half-width of the maximum of real part of the incomplete gamma function entering $c_2$, as a function of $P^2$, at $\mu=0$,
\begin{equation}\label{half-width}
    \De=2\sqrt{\frac{\epsilon}{|\la|}}e^{-\ga_\mathrm{E}/2},
\end{equation}
where $\ga_\mathrm{E}$ is the Euler constant. We are interested in the real part of the contribution because the poles appear by complex conjugate pairs in the expansion \eqref{expansion} such that the imaginary part is canceled out. To chop off the exponentially growing terms (at small $P^2$), we additionally assume
\begin{equation}\label{cond_la_2}
    |\la|\leq\chi/b,\qquad\chi\approx0.4.
\end{equation}
The last number is taken directly from the plot of the incomplete gamma function at $\nu=0$. If the conditions \eqref{cond_la_1} and \eqref{cond_la_2} are fulfilled then we have to retain only the first several terms in the sum over $q$ which are near to the point $q=0$. As for the sum over $p$, we should keep only such terms that
\begin{equation}\label{P_range}
    P^2\in(b-\De/2,b+\De/2).
\end{equation}
If the metric $g_{ij}$ is approximately isotropic (all of its eigenvalues are of the order of $q_m^2$) and $\epsilon$ is small then the expression for $c_{os}$ contains only a few number of relevant terms.

The case of a strongly anisotropic metric needs a more cumbersome procedure. In this case, an immediate application of the above formulas is not very effective since there are a lot of terms which satisfy the condition \eqref{P_range}. These are the terms that correspond to the vectors $P_i$ differing by the components along the large principal axes of the metric $g_{ij}$. This problem is also easy to see from the expression \eqref{contrib_p_one_pole} when $|\la|$ is of the order of the least eigenvalue $q_m^2$ of the metric.

Now we show how to modify the above procedure to the case of an isotropic metric. Let the eigenvalues of the metric $g_{ij}$ constitute a hierarchy
\begin{equation}\label{eigen_hierarchy}
    \bar{\la}_1\ll\bar{\la}_2\ll\cdots\ll\bar{\la}_n.
\end{equation}
For brevity, we shall suppose that the metric is diagonal. Then, we part the integral \eqref{contrib_one_pole_c} according to this hierarchy
\begin{equation}\label{integral_hier}
    c=-\pi^{d/2}\sqrt{g}\sum_q\left[\int_0^{\la_1}idt+\int_{\la_1}^{\la_2}idt+\ldots+\int_{\la_n}^\infty idt\right](it)^{-1-d/2}e^{i(\pi^2 q^2/t+s_0 t+2\pi a_iq^i)},
\end{equation}
where $\la_i$ are complex numbers with moduli being of the order of the respective eigenvalues $\bar{\la}_i$. As before, we assume that $\la_i=-i|\la_i|$. We have already known the first and the last integrals in \eqref{integral_hier}. It remains to consider a contribution of the $k$-th integral
\begin{equation}
    c_k:=-\pi^{d/2}\sqrt{g}\sum_q\int_{\la_{k-1}}^{\la_k}idt(it)^{-1-d/2}e^{i(\pi^2 q^2/t+s_0 t+2\pi a_iq^i)}.
\end{equation}
It is useful to denote by $q^a$  all the indices $q$ corresponding to the subspace of the metric eigenvectors with the eigenvalues  $\bar{\la}_i\leq\bar{\la}_{k-1}$, and by $q^A$ all the rest indices $q$. We also denote by $R_k$ a number of indices $q^a$ and assign by definition  $R_1:=0$. Then
\begin{equation}\label{c_k}
    c_k=-\pi^{(d-R_k)/2}\sqrt{\frac{g}{g_a}}\sum_{p_a,q^A}\int_{|\la_{k-1}|}^{|\la_k|}dtt^{-1-(d-R_k)/2}e^{-\pi^2 q^2_A/t-(P_a^2-s_0)t-2\pi ia_Aq^A},
\end{equation}
where $g_a:=\det g_{ab}$, $q_A^2:=g_{AB}q^Aq^B$, and so on. Taking into account the range of variable $t$, it is now easy to see that the sum in the last formula can be terminated with a neglible error. A contribution of the zero mode has the form
\begin{multline}
    c_k^0=-\pi^{(d-R_k)/2}\sqrt{\frac{g}{g_a}}\sum_{p_a}(P_a^2-s_0)^{(d-R_k)/2}\\
    \times\left[\Ga\left((R_k-d)/2,|\la_{k-1}|(P_a^2-s_0)\right)-\Ga\left((R_k-d)/2,|\la_{k}|(P_a^2-s_0)\right)\right].
\end{multline}
This expression is a monotonically decreasing function of $P^2_a$. With an increase of the indices $P_a$, it tends exponentially to zero. Hence, we can retain only the first leading terms in this sum for which $P^2_a\approx0$.

An asymptotic expansion of the integral at $q^A\neq0$ can be obtained by the WKB-method (see, e.g., \cite{Fedoryuk}). This method leads to the five cases corresponding to different values of the indices $q^A$ and $P_a$. They are
\begin{multline}\label{c_k_tilde}
    \tilde{c}_k\approx-\sqrt{\frac{g}{g_a}}\sideset{}{'}\sum_{q^A}\sum_{p_a}e^{-2\pi ia_Aq^A}\times\\
\left\{
                                \begin{array}{ll}
                                  (\pi/|\la_k|)^{(d-R_k)/2}\sum_{n=0}^\infty\left[\left(\frac{-1}{S'_1(t)}\frac{d}{dt}\right)^n\frac{t^{-1-(d-R_k)/2}}{S'_1(t)}\right]_{t=1}e^{S_1(1)}, & \pi^2q_A^2/|\la_k|+|\la_k|(b-P^2_a)\gtrsim1; \\
                                  f_{(d-R_k)/2}(q_A,\sqrt{P^2_a-s_0}), & \pi^2q_A^2/|\la_k|+|\la_k|(b-P^2_a)\approx0; \\
                                  2k_{(d-R_k)/2}(q_A,\sqrt{P^2_a-s_0}), & |\la_{k-1}|\lesssim\pi q_A/\sqrt{P^2_a-b}\lesssim|\la_k|; \\
                                  f_{(R_k-d)/2}(\sqrt{P^2_a-s_0},q_A), & \pi^2q_A^2/|\la_{k-1}|+|\la_{k-1}|(b-P^2_a)\approx0; \\
                                  -(\pi/|\la_{k-1}|)^{(d-R_k)/2}\sum_{n=0}^\infty\left[\left(\frac{-1}{S'_2(t)}\frac{d}{dt}\right)^n\frac{t^{-1-(d-R_k)/2}}{S'_2(t)}\right]_{t=1}e^{S_2(1)}, & \pi^2q_A^2/|\la_{k-1}|+|\la_{k-1}|(b-P^2_a)\lesssim-1;
                                \end{array}
                              \right.
\end{multline}
where $k_\nu(q,p)$ is the Macdonald function defined in Eq. \eqref{Bessel_func}. We also introduce the notation
\begin{equation}
\begin{gathered}
    S_1(t)=-\frac{\pi^2q^2_A}{|\la_k|t}+|\la_k|(s_0-P_a^2)t,\qquad S_2(t)=-\frac{\pi^2q^2_A}{|\la_{k-1}|t}+|\la_{k-1}|(s_0-P_a^2)t,\\
    f_\nu(q,p)=\int_0^{q/p}dt t^{-1-\nu}e^{-\pi(q^2/t+p^2t)}.
\end{gathered}
\end{equation}
The function $f_\nu(q,p)$ is not a cylinder function and apparently cannot be expressed in a simple form (see the remark on p. 313 in \cite{Wats}). One can think of this function as a ``half'' of the Macdonald function in the sense that
\begin{equation}
    f_\nu(q,p)+f_{-\nu}(p,q)=2k_\nu(q,p).
\end{equation}
At large arguments, the function $f_\nu(q,p)$ possesses the asymptotic expansion analogous to \eqref{Hankel_func_expans}
\begin{equation}
    f_\nu(q,p)\approx\left(\frac{p}{q}\right)^\nu e^{-2\pi qp}\sum_{n=0}^\infty\sum_{s=0}^\infty\sum_{k=0}^n\frac{2^n(-1)^{n-k} C_n^k}{(4\pi qp)^{(n+k+1)/2+s}}\frac{\Ga(1-\nu)\Ga((n-k+1)/2)\Ga((n+k+1)/2+s)}{n!s!\Ga(1-\nu-n)\Ga((n-k+1)/2-s)}.
\end{equation}
In increasing $P^2_a$ with fixed $q_a^2$, the above mentioned cases are realized successively from top to down. It is clear from the representation of the integral \eqref{c_k} that its value rapidly decreases when $P^2_a$ increases. One can see this from the explicit expressions \eqref{c_k_tilde} as well. In other words, to obtain an estimation of the contribution of this integral to the $\Omega$-potential, it will be sufficient to choose the first case (at the given value of $b$) from the indicated ones and take a few first values of the indices $q^A$ and $p_a$.

In summary, the contribution to the oscillating part of the partition function from a pole approaching the real axis reads as
\begin{equation}
   c_{os}=\sum_{i=1}^n(\tilde{c}_i+c^0_i)+c_\infty+\pi^{d/2}\sqrt{g}(-s_0)^{d/2}\Ga(-d/2),
\end{equation}
where we should take $\la$ to be equal to $\la_n$ in the expression for $c_\infty$. The value of $\la_n$ needs to meet the condition  \eqref{cond_la_2}. Otherwise, in the sum \eqref{contrib_p_one_pole}, we have to take into account the contributions not only from the indices $p_i$ satisfying \eqref{P_range}, but also from that indices $p_i$ for which $P^2$ is small.

\subsection{Quasiclassical contribution}\label{Quasi_Contr}

\subsubsection{Nonrelativistic limit}

Let us turn to the zero mode $q=0$ of the integral $I_d^q$ (quasiclassical contribution). On integrating over angles, it can be cast into the form
\begin{equation}\label{Id0}
    I_d^0=\frac{2\pi^{d/2}\sqrt{g}}{\Gamma{\left(d/2\right)}}\int_0^\infty d p p^{d-1}\ln(1+ e^{\mu-\omega(p^2+m^2)}).
\end{equation}
Now we introduce a useful notation
\begin{equation}\label{S}
    \omega(p^2+m^2)-\omega_0=:y(s),\qquad\omega_0:=\omega(m^2),\qquad\bar{\omega}_0:=\omega_0-\mu,
\end{equation}
where $s:=p^2$. In order to get rid of the logarithm in the integral \eqref{Id0}, we integrate it by parts and substitute a series representation of $s^{\alpha}(y)$ to it,
\begin{equation}\label{y(x)}
    s^{\alpha}(y)= \sum\limits_{k=0}^{\infty}\lambda_k^{\alpha}y^{\alpha+k}, \qquad \lambda_k^{\alpha}=\lim_{s\rightarrow0}\frac{1}{k!}\left[\frac{1}{y'(s)}\frac{d}{d s}\right]^k \left[\frac{s}{y(s)}\right]^{\alpha}.
\end{equation}
Making use of the definition of the polylogarithm function \eqref{polylog def}, we arrive at the expansion (for the relativistic dispersion law see, e.g., \cite{polylog})
\begin{equation}\label{ID0-true}
    I_d^0=-\pi^{d/2}\sqrt{g}\sum\limits_{k=0}^{\infty}\frac{\Gamma{\left(d/2+1+k\right)}}{\Gamma{\left(d/2+1\right)}}
    \lambda_k^ { d/2 }\Li_{d/2+1+k}\left(-e^{-\bar{\omega}_0}\right).
\end{equation}
The analogous expansion for the Bose-Einstein distribution is recovered by a change of the overall sign and the sign in the argument of the polylogarithm. Since $y$ is proportional to $\beta$, the expansion \eqref{ID0-true} is carried out in inverse powers of $\beta$ and provides the low-temperature asymptotics of the quasiclassical contribution to the $\Omega$-potential. In particular, it terminates on the first term for the nonrelativistic dispersion law.

\subsubsection{High-temperature expansion}

In order to obtain the high-temperature expansion, we shall proceed in the analogous way. On integrating by parts in the integral \eqref{Id0}, we have for different statistics
\begin{equation}\label{Id0_hightem}
    I_d^0=\frac{\pi^{d/2}\sqrt{g}}{\Gamma(d/2+1)}\int_{\beta\bar{\omega}_0}^\infty dy\frac{\left(\omega^{-1}(\beta^{-1}y+\mu)-m^2\right)^{d/2}}{e^y\pm1},
\end{equation}
where we have shown explicitly a dependence on $\beta$ and supposed that $\omega(s)$ is a monotonically increasing function. Our aim is to expand this integral in increasing powers of $\beta$. First, we expand a numerator of fraction entering the integrand. Assume that, at sufficiently large $|y|$, the following representation holds
\begin{equation}\label{omega_inv_expans}
    [\omega^{-1}(y)]^{\alpha}=\sum_{p=0}^\infty\ka_p^\al y^{\ga\al-p},\qquad\ka^\al_p=\lim_{s\rightarrow0}\frac{(-1)^p}{p!}\left[\frac{\omega^2(s)}{\omega'(s)}\frac{d}{ds}\right]^p\left[\frac{s}{\omega^\ga(s)}\right]^\al,
\end{equation}
where $\ga>0$ and $\omega^{-1}(y)$ is an analytic function of $\ga$. Then, sequentially expanding the expression in series, we obtain
\begin{multline}\label{expansion high temp}
    \left(\omega^{-1}(\beta^{-1}y+\mu)-m^2\right)^{d/2}=\sum_{k,p,l=0}^\infty\frac{(-1)^k}{k!l!}\frac{\ka_p^{d/2-k}\Gamma(d/2+1)\Ga(\ga(d/2-k)-p+1)}{\Ga(d/2+1-k)\Ga(\ga(d/2-k)-l-p+1)}\\
    \times m^{2k}\mu^l(\beta^{-1}y)^{\ga(d/2-k)-l-p}.
\end{multline}
This power series converges at $|y|>\beta R$ at sufficiently large $R$ independent of $\beta$.

Further, we need to use the well-known expansion of the incomplete $\zeta$-function
\begin{equation}\label{incompl_zeta}
\begin{split}
    \int_a^\infty dx\frac{x^{\nu-1}}{e^x-1}&=\Gamma(\nu)\zeta(\nu)-\sum_{n=-1}^\infty\frac{(-1)^n\zeta(-n)a^{\nu+n}}{\Gamma(n+1)(\nu+n)},\\ \int_a^\infty dx\frac{x^{\nu-1}}{e^x+1}&=(1-2^{1-\nu})\Gamma(\nu)\zeta(\nu)-\sum_{n=0}^\infty(1-2^{1+n})\frac{(-1)^n\zeta(-n)a^{\nu+n}}{\Gamma(n+1)(\nu+n)}.
\end{split}
\end{equation}
In the bosonic case, this expansion is valid at $|a|<2\pi$, while, in the fermionic case, it holds at $|a|<\pi$. The functions in the RHS of the equalities are entire functions of $\nu$. Let us part the integral \eqref{Id0_hightem} into two: $[\beta\bar{\omega}_0,\beta R]$ and $[\beta R,+\infty)$; and denote these integrals as $i_1$ and $i_2$, respectively. Then, at sufficiently small $\beta$, we can apply formulas \eqref{incompl_zeta} to the second integral $i_2$. Substituting the expansion \eqref{expansion high temp} into $i_2$, we integrate the series term by term. As a result, the whole expansion of the integral $i_2$ splits naturally into two contributions from the first and second terms in Eqs. \eqref{incompl_zeta}, respectively. In the first contribution to $i_2$, the expansion is carried out in increasing powers of $\beta$, there being a finite number of terms at any fixed power of $\beta$. Indeed, in the bosonic case
\begin{equation}\label{Id0_hightem_fin}
    i_2=\pi^{d/2}\sqrt{g}\sum_{k,l,p=0}^\infty \frac{(-1)^k\ka_p^{d/2-k}m^{2k}\mu^l\beta^{\ga(k-d/2)+l+p}}{k!l!\Ga(d/2+1-k)}\Gamma(\ga(d/2-k)-p+1)\zeta(\ga(d/2-k)-l-p+1)+\vf(\beta R),
\end{equation}
where $\vf(\beta R)$ denotes the second contribution. We also explicitly marks its dependence on the lower integration limit in \eqref{Id0_hightem}. This last term has to be resummed. From Eq. \eqref{incompl_zeta} it is not difficult to see that every term of the series in $n$ in the expression for $\vf(\beta R)$ can be written in the integral form
\begin{equation}\label{i_2}
    \frac{\pi^{d/2}\sqrt{g}}{\Gamma(d/2+1)}\frac{(-1)^n\zeta(-n)}{\Gamma(n+1)}\int_{\beta R}^\infty dyy^n \left(\omega^{-1}(\beta^{-1}y+\mu)-m^2\right)^{d/2},
\end{equation}
where $d$ and $\ga$ are chosen so that the integral converges and then continued by analyticity to their initial values. Bearing in mind the Taylor expansion of the function $(e^y-1)^{-1}$ in $y$ in a vicinity of zero, the expression for the integral $i_1$ can be cast into the form \eqref{i_2}. Matching the parameters $d$ and $\ga$ in the both integrals, we can combine them into one. This gives
\begin{equation}\label{phi_part}
    i_1+\vf(\beta R)=\pi^{d/2}\sqrt{g}\sum_{n=-1}^\infty\frac{(-1)^n\zeta(-n)}{\Gamma(n+1)}\beta^{n+1}\int_{\omega_0}^\infty dy(y-\mu)^n\frac{\left(\omega^{-1}(y)-m^2\right)^{d/2}}{\Ga(d/2+1)}.
\end{equation}
For the Fermi-Dirac distribution, the series \eqref{Id0_hightem_fin} and \eqref{phi_part} have the same form as in the bosonic case. The difference consists in the overall sign, powers of two typical for the fermionic distribution and other summation limits: $n=\overline{0,\infty}$ (cf. Eqs. \eqref{incompl_zeta}). Thus, in both cases, the problem reduces to a finding of the integral
\begin{equation}\label{sigma-n}
    m^d\s_d^n(m^2,\mu)=\int_{\omega_0}^\infty dy(y-\mu)^n\frac{\left(\omega^{-1}(y)-m^2\right)^{d/2}}{\Ga(d/2+1)}=\int_0^\infty \frac{dss^{d/2}}{\Ga(d/2+1)}\omega'(s+m^2)\left(\omega(s+m^2)-\mu\right)^n,
\end{equation}
in the form of an analytic function of the complex variables $d$ and $\ga$. It is clear that this integral converges at
\begin{equation}
    \re d>-2,\qquad\re(\ga d)<-2n-2.
\end{equation}
The factor $m^d$ is introduced to Eq. \eqref{sigma-n} just for convenience.

If the parameters $m$ and $\mu$ are arbitrary, the expression \eqref{sigma-n} for $\s_d^n$ does not contain any small parameter and cannot be evaluated or expanded. Therefore, we need to specify a form of the dispersion law $\omega(s)$ in order to proceed to a further investigation. For example, for a homogeneous dispersion law,
\begin{equation}
    \omega=s^{1/\ga},\qquad\ka^\al_p=\de^0_p,
\end{equation}
the sum over $p$ in the expression \eqref{Id0_hightem_fin} stops at the first term and the integral \eqref{sigma-n} reduces to
\begin{multline}
    \s_d^n=m^{2(n+1)/\ga}\int_1^\infty dx(x-z)^n\frac{(x^\ga-1)^{d/2}}{\Gamma(d/2+1)}\\
    =m^{2(n+1)/\ga}\ga^{-1}\sum_{s=0}^\infty\frac{\Ga((s-n-1)/\ga-d/2)}{\Ga((s-n-1)/\ga+1)}\frac{\Ga(n+1)(-z)^{s}}{s!\Ga(n-s+1)},
\end{multline}
where $z=\mu m^{-2/\ga}$. This series representation is valid at $|z|<1$ and also at the nonnegative integer $n$ when the series terminates. The high-temperature expansion for the Fermi-Dirac distribution depends on $\s_d^n$ with nonnegative integer $n$ only. So, in the fermionic case, the high-temperature expansion of the integral \eqref{Id0_hightem} reads as
\begin{multline}\label{high_temp_ferm}
    I_d^0=\pi^{d/2}\sqrt{g}\sum_{l=0}^\infty\biggl[\sum_{k=0}^\infty (1-2^{l-\ga(d/2-k)})\frac{(-m^2)^k\mu^l\beta^{\ga(k-d/2)+l}}{k!l!\Ga(d/2+1-k)}\Gamma(\ga(d/2-k)+1)\zeta(\ga(d/2-k)-l+1)\\
    +\ga^{-1}m^d(1-2^{1+l})\zeta(-l)(\beta m^{2/\ga})^{l+1}\sum_{k=0}^l\frac{\Ga(-d/2-(k+1)/\ga)}{\Ga(1-(k+1)/\ga)}\frac{(-1)^kz^{l-k}}{k!(l-k)!}\biggr].
\end{multline}
Note that a ``half'' of the contributions from the second term in the square brackets vanish due to the property of the $\zeta$-function. These are the terms at the even positive $l$.

In order to obtain the high-temperature expansion for the Bose-Einstein distribution law, we need to evaluate the integral
\begin{equation}\label{sigma-1_gen}
    \s_d^{-1}(z)=\int_1^\infty \frac{dx}{\Gamma(d/2+1)}\frac{(x^\ga-1)^{d/2}}{x-z}=\ga^{-1}\sum_{l=0}^\infty\frac{\Ga(l/\ga-d/2)}{\Ga(l/\ga+1)}z^l.
\end{equation}
This integral is an analytic function of $z$ with a branch cut discontinuity along the part of the real axis $z>1$. The discontinuity on the cut is equal to
\begin{equation}
    \s_d^{-1}(z+i\epsilon)-\s_d^{-1}(z-i\epsilon)=2\pi i\frac{(z^\ga-1)^{d/2}}{\Ga(d/2+1)}.
\end{equation}
The above series representation holds for $|z|<1$. It allows us to obtain the analytic continuation of this integral to arbitrary complex values of $d$ and $\ga$. If $\ga$ is a natural number then this integral can be written as a finite sum of the Gauss hypergeometric functions
\begin{equation}
    \s_d^{-1}(z)=\ga^{-1}\sum_{n=0}^{\ga-1}\frac{\Ga(n/\ga-d/2)}{\Ga(n/\ga+1)}z^nF(1,n/\ga-d/2;n/\ga+1;z^\ga).
\end{equation}
At $\ga=2$ (relativistic dispersion law), we have different equivalent representations
\begin{multline}\label{sigma-1}
    \s_d^{-1}(z)=\frac{2^d\Ga(-d)}{\Ga(1-d/2)}F(1,-d;1-d/2;(1+z)/2)=-\frac{\Ga\left((1-d)/2\right)}{d\sqrt\pi}F(1/2,-d/2;1-d/2;1-z^2)\\
    =\frac12\Ga(-d/2)(1-z^2)^{d/2}+\Ga\left((1-d)/2\right)\frac{z}{\sqrt{\pi}}F(1,(1-d)/2;3/2;z^2)=2^{d/2}\Ga(-d)(1-z^2)^{d/4}\mathrm{P}^{d/2}_{d/2}(-z),
\end{multline}
where $\mathrm{P}^\mu_\nu(z)$ is an associated Legendre function of the first kind \cite{BatErde} single-valued and regular at $|z|<1$. For an arbitrary $\ga$, it is easy to obtain the following recurrence relation
\begin{equation}
    \ga z^{-\ga(d/2+1)-1}\s_d^{-1}(z)=\partial_z[z^{-\ga(d/2+1)}\s^{-1}_{d+2}(z)].
\end{equation}
To derive the expansion of $\s_d^{-1}(z)$ at $|z|>1$, we write it in the equivalent form
\begin{equation}
    \Ga(d/2+1)\s_d^{-1}(z)=\int_0^\infty d\al\int_1^\infty dx(x^\ga-1)^{d/2}e^{-\al(x-z)},\qquad\re z<0.
\end{equation}
Now we expand the preexponential factor in $x$ and integrate the series term by term. Then, by the use of the expansion for the incomplete gamma function \eqref{Gamma_func_inco}, we expand the obtained preexponential factor in $\al$ and again integrate the series term by term. Thereby, we come to
\begin{equation}
    \s_d^{-1}(z)=\sum_{k=0}^\infty\frac{(-1)^k}{k!\Ga(d/2+1-k)}\left[\frac{\pi(-z)^{\ga(d/2-k)}}{\sin\pi\ga(k-d/2)}+\sum_{n=0}^\infty\frac{z^{-1-n}}{\ga(d/2-k)+n+1}\right].
\end{equation}
The series in $k$ in the last term in the square brackets is easily summed and expressed in terms of the beta function. As a result, we have
\begin{equation}
    \s_d^{-1}(z)=\sum_{k=0}^\infty\left[\frac{(-1)^k}{k!\Ga(d/2+1-k)}\frac{\pi(-z)^{\ga(d/2-k)}}{\sin\pi\ga(k-d/2)}-\ga^{-1}\frac{\Ga(-d/2-(k+1)/\ga)}{\Ga(1-(k+1)/\ga)}z^{-1-k}\right].
\end{equation}
If $\ga$ is a natural number, we deduce the representation
\begin{equation}
    \s_d^{-1}(z)=-\frac{\pi(-z)^{\ga d/2}}{\sin(\pi\ga d/2)}\frac{(1-z^{-\ga})^{d/2}}{\Ga(d/2+1)}-\ga^{-1}\sum_{k=1}^\ga\frac{\Ga(-d/2-k/\ga)}{\Ga(1-k/\ga)}z^{-k}F(1,k/\ga;k/\ga+d/2+1;z^{-\ga}).
\end{equation}

Thus the high-temperature expansion in the bosonic case becomes
\begin{multline}\label{Id0_hightem_fin_hom}
    I_d^0=\pi^{d/2}\sqrt{g}\sum_{l=0}^\infty\biggl[\sum_{k=0}^\infty\frac{(-m^2)^k\mu^l\beta^{\ga(k-d/2)+l}}{k!l!\Gamma(d/2+1-k)}\Gamma(\ga(d/2-k)+1)\zeta(\ga(d/2-k)-l+1)+\\
    +\ga^{-1}m^d(\beta m^{2/\ga})^{l+1}\zeta(-l)\sum_{k=0}^l\frac{(-1)^kz^{l-k}}{k!(l-k)!}\frac{\Ga(-d/2-(k+1)/\ga)}{\Ga(1-(k+1)/\ga)}\biggr]+\pi^{d/2}\sqrt{g}m^d\s_d^{-1}(z).
\end{multline}
It is clear that the obtained high-temperature expansions \eqref{high_temp_ferm} and \eqref{Id0_hightem_fin_hom} are straightforwardly generalized to the dispersion laws in the form of a finite sum of homogeneous functions. In the expression \eqref{Id0_hightem_fin_hom}, as well as in the high-temperature expansion for the Fermi-Dirac distribution \eqref{high_temp_ferm}, we can set $d$ to be equal to the physical dimension and, henceforward, we shall denote the physical dimension as $d$. Then the two terms in the brackets and the last term possess the poles in the $\ga$-plane which are mutually canceled out.

In the fermionic case, the singularities arise when the physical degree of homogeneity $\bar{\ga}$ of the dispersion law is a rational number of the form
\begin{equation}\label{gamma_distinguish}
\begin{aligned}
    \bar{\ga}&=\frac{2p+2}{2q+1},&\quad\text{for odd $d$};\\
    \bar{\ga}&=\frac{p+1}{q+1},&\quad\text{for even $d$};
\end{aligned}
\end{equation}
where $p$ and $q$ are nonnegative integers. Denoting by $l_1$ and $k_1$ the indices in the first term of the high-temperature expansion \eqref{high_temp_ferm} and by $l_2$ and $k_2$ the indices in the second term, we see that the cancelation of singularities occurs between the terms with
\begin{equation}
    l_1=l_2-k_2,\quad k_2\leq l_2;\qquad k_1-\frac{d}2=\frac{k_2+1}{\bar{\ga}},
\end{equation}
at
\begin{equation}
\begin{aligned}
    (k_1,k_2)&=\left(\frac{d+1}{2}+q+n(2q+1),p+n(2p+2)\right),&\quad n&=\overline{0,\infty},&\qquad\text{for odd $d$};\\
    (k_1,k_2)&=\left(\frac{d}{2}+n(q+1),n(p+1)-1\right),&\quad n&=\overline{1,\infty},&\qquad\text{for even $d$}.
\end{aligned}
\end{equation}
In the bosonic case, there are extra singularities coming from the pole of the $\zeta$-function in unity. They are canceled out by the singularities of the last term in \eqref{Id0_hightem_fin_hom}. Expanding this last term in the series \eqref{sigma-1_gen} and denoting the corresponding summation index as $l_3$, it is not difficult to understand that the cancelation occurs when
\begin{equation}\label{singul_type_II}
    l_1=l_3;\qquad l_1=\bar{\ga}(d/2-k_1),\quad k_1=\overline{0,[d/2]},
\end{equation}
that is, when $\bar{\ga}$ is of the form \eqref{gamma_distinguish} and
\begin{equation}
\begin{aligned}
    (k_1,l_1)&=\left(\frac{d-1}{2}-q-n(2q+1),(2n+1)(p+1)\right),&\quad n=&\overline{0,\left[\frac{d}{4q+2}-\frac12\right]},&\qquad\text{for odd $d$};\\
    (k_1,l_1)&=\left(\frac{d}{2}-n(q+1),n(p+1)\right),&\quad n=&\overline{0,\left[\frac{d}{2(q+1)}\right]},&\qquad\text{for even $d$}.
\end{aligned}
\end{equation}
Note that the terms proportional to $\ln(\beta m^{2/\bar{\ga}})$ may appear in the high-temperature expansion for homogenous dispersion laws only with the degree of homogeneity of the form \eqref{gamma_distinguish}. Since only in this case we need to expand the powers of $\beta$ and $m$ in the expressions \eqref{high_temp_ferm} or \eqref{Id0_hightem_fin_hom} into series in $\ga$.

As an example of the cancelation of singularities in \eqref{Id0_hightem_fin_hom}, we consider the bosonic high-temperature expansion with the natural degree of homogeneity $\bar{\ga}\in\mathbb{N}$. Besides, we are only interested in the singularities of the type \eqref{singul_type_II} when, upon the cancelation of singularities, we should sum an infinite series in $z$ to obtain a closed-form expression. Then
\begin{multline}\label{high_temp_bos}
    I_d^0=\pi^{d/2}\sqrt{g}\biggl[\sideset{}{'}\sum_{l,k=0}^\infty\frac{(-m^2)^k\mu^l\beta^{\ga(k-d/2)+l}}{k!l!\Gamma(d/2+1-k)}\Gamma(\ga(d/2-k)+1)\zeta(\ga(d/2-k)-l+1)+\\
    +\ga^{-1}m^d\sum_{l=0}^\infty(\beta  m^{2/\ga})^{l+1}\zeta(-l)\sum_{k=0}^l\frac{(-1)^kz^{l-k}}{k!(l-k)!}\frac{\Ga(-d/2-(k+1)/\ga)}{\Ga(1-(k+1)/\ga)}\biggr]_{\ga\rightarrow\bar{\ga}}\\
    +\pi^{d/2}\sqrt{g}m^d\bar{\ga}^{-1}\biggl[\sideset{}{'}\sum_{s=1}^{\bar{\ga} d/2-1}\frac{\Ga(s/\bar{\ga}-d/2)}{\Ga(s/\bar{\ga}+1)}z^s
    +\sum_{k=1}^{\bar{\ga}}\frac{\Ga(k/\bar{\ga})z^{\bar{\ga}d/2+k}}{\Ga(k/\bar{\ga}+d/2+1)}F(1,k/\bar{\ga};k/\bar{\ga}+d/2+1;z^{\bar{\ga}})\biggr]\\
    -\pi^{d/2}\sqrt{g}\sum_{k=0}^{[d/2]}\frac{(-m^2)^k\mu^{\bar{\ga}(d/2-k)}}{k!\Ga(d/2+1-k)}\Bigl[\ln\beta m^{2/\bar{\ga}}-\left(\ga_{\mathrm{E}}+\psi\left(\bar{\ga}(d/2-k)\right)\right)+\bar{\ga}^{-1}(\psi(d/2+1-k)-\psi(k+1))\Bigr],
\end{multline}
where $\psi(x)=\Gamma'(x)/\Gamma(x)$, the primed sum over $l$ denotes a summation over those $l$ at which the argument of the $\zeta$-function is not unity, whereas the prime at the sum over $s$ says that we throw away the singular terms arising from a nonpositive integer value of the argument of the gamma function. In the odd-dimensional space, the above high-temperature expansion holds only for even $\bar{\ga}$ where there exist  singularities of the type \eqref{singul_type_II}. Specific values of the Gauss hypergeometric function can be found, for example, in \cite{PrBrMaIII}. In the case of relativistic and nonrelativistic spectra, all the hypergeometric functions entering the last formula are expressed in terms of elementary functions.

In conclusion of this section, we write out the linear in $\beta$ terms of the high-temperature expansion for the fermionic $\Omega$-potential at a vanishing chemical potential. As we know, these terms are of a particular interest since they have the same form as a finite part of the one-loop contribution to the vacuum energy \eqref{zero_point_energ}. Starting from the general formula \eqref{high_temp_ferm}, it is not difficult to find this contribution. Tending $\ga$ to its physical value $\bar{\ga}$, we have
\begin{equation}
    I^0_d=\pi^{d/2}\sqrt{g}\beta m^{d+2/\bar{\ga}}\frac{\Ga(-d/2-1/\bar{\ga})}{2\bar{\ga}\Ga(1-1/\bar{\ga})}+\ldots
\end{equation}
when $d/2+1/\bar{\ga}$ is not natural,
\begin{equation}
    I^0_d=\pi^{d/2}\sqrt{g}\beta m^{d+2/\bar{\ga}}\frac{(-1)^{d/2+1}\Ga(1+1/\bar{\ga})}{2\Ga(d/2+1/\bar{\ga}+1)}+\ldots
\end{equation}
when $d$ is even, but $\bar{\ga}^{-1}$ is natural, and
\begin{multline}
    I^0_d=\frac{\pi^{d/2}\sqrt{g}(-1)^{d/2+1/\bar{\ga}+1}\beta m^{d+2/\bar{\ga}}}{2\Ga(1-1/\bar{\ga})\Ga(d/2+1/\bar{\ga}+1)}\\
    \times\left\{\ln\frac{\beta m^{2/\bar{\ga}}}{ 2^{2/\bar{\ga}-1}\pi}+\ga_{\mathrm{E}}+\bar{\ga}^{-1}[2\psi(2-2/\bar{\ga})-\psi(3/2-1/\bar{\ga})-\psi(d/2+1/\bar{\ga}+1)]\right\}+\ldots
\end{multline}
when $d$ is odd and $\bar{\ga}^{-1}$ is half-integer. The last formula gives, in particular, a finite part of the quasiclassical contribution to the vacuum energy for particles with the relativistic dispersion law in a three-dimensional space.

\section{Examples}\label{Examp}

\subsection{Massless particles}\label{Massl_Part}

\subsubsection{Zero chemical potential}

In this section, in order to illustrate the main features of the developed general procedure, we consider a simple model of a gas of free massless particles confined to the rectangular parallelepiped with sizes $(L_x,L_y,L_z)$ both at zero and non-zero chemical potential. We start with the case of a vanishing chemical potential. Such a model describes, for instance, a gas of photons in an ideal metal rectangular box. Notwithstanding this model is profoundly investigated (see \cite{CasPold,AmbWolf} and some recent papers \cite{HJKS,FKKLM,Marach,LimTeo,LimTeo1,ElOdSah,LCLZ,CNSS,Edery,MREGJK,JVH,GeKlMo}), its consideration will be instructive because of its high degeneracy. We shall derive new rapidly convergent expansions for the $\Omega$-potential applying the resummation formulas found in Sec. \ref{Resum}. For the sake of conciseness, we shall study the fields subjected to the periodic and Dirichlet boundary conditions. For the model at issue, the partition function corresponding to the quantum fields with the Neumann boundary conditions can be reduced to the partition function of the fields with the periodic boundary conditions (see, e.g., \cite{LimTeo,Edery}).

The ``metric'' is defined in this case as
\begin{equation}
    g_{ij}=l_T^{-2}diag(L^2_x,L^2_y,L^2_z),\qquad l_T=\pi\beta c\hbar,
\end{equation}
where $l_T$ is a  thermal wave-length of a massless particle. This quantity is a characteristic scale of the model. At the temperature $\beta^{-1}=1$ K, it approximately equals to $0.72$ cm. The roots of Eq. \eqref{zeros} take a simple form
\begin{equation}\label{roots rel msls}
    p_k=\mu+i\omega_k,\qquad p^*_k=-\mu+i\omega_k,\qquad k\geq0.
\end{equation}
As we have already noted, the poles \eqref{roots rel msls} lie on the ``physical'' sheet of the function $\omega(s)$ only if $\mu>0$. At the vanishing chemical potential, they contribute to the oscillating part of the $\Omega$-potential with the factor $1/2$, the integral over the cut being understood in the sense of principal value. Further, we shall assume that $\mu=0$, but, in some formulas, it will be useful to put $\mu$ to zero only in the finial result.

Let us first suppose that a summation over the quantum numbers $p_i$ (do not confuse with the roots $p_k$ of Eq. \eqref{roots rel msls}) is carried over the infinite ranges $(-\infty,\infty)$. Then the oscillating contribution to the logarithm of the partition function in $d$ dimensions reads as
\begin{equation}
    \pm\beta\Omega^{os}_d=-i\pi\sqrt{g}\sideset{}{'}\sum_q\sum_{k=0}^\infty h^{(1)}_{d/2}(q,i\omega_k),
\end{equation}
where minus corresponds to bosons and plus is to be taken for fermions. In the bosonic case, the term at $k=0$ comes with the factor $1/2$. As long as $\re p_k=0$, the oscillating contribution does not, in fact, oscillate since the Hankel functions turn into the Macdonald ones with real arguments. However, we shall call it oscillating to distinguish from the other contributions to the $\Omega$-potential. By the use of the asymptotic expansion of the Hankel function \eqref{Hankel_func_expans}, we can sum the series in $k$. In the fermionic case, it becomes
\begin{multline}
    \beta\Omega^{os}_d\approx\frac{\sqrt{g}}{2\pi}\sideset{}{'}\sum_q\left(\frac{2\pi}{q}\right)^{(d+1)/2}\sum_{s=0}^\infty\frac{(8\pi^2q)^{-s}\Ga\left((d+1)/2+s\right)}{s!\Ga\left((d+1)/2-s\right)}\\
    \times\left[\Li_{s-(d-1)/2}\left(e^{-4\pi^2q}\right)-2^{s-(d-1)/2}\Li_{s-(d-1)/2}\left(e^{-2\pi^2q}\right)\right].
\end{multline}
If the condition \eqref{cond rel_high} is satisfied then the series in $q$ can be broken off taking into account the first several terms only. For odd $d$, the series in $s$ terminates, all the polylogarithms entering the formula are expressed in terms of elementary functions, and we obtain the exact expression for the $\Omega$-potential. As for even $d$, the series in $s$ can be also terminated with a neglible error. The optimal number $s$, at which the series should be broken off, is the number of the series term with a minimal absolute value (see the remark after Eq. \eqref{Hankel_func_expans}). The oscillating contribution becomes significant only when the characteristic scale of the system is of the order of $l_T$ or smaller. On the scale $l_T$, a relative contribution of the oscillating term grows exponentially with decreasing sizes of the system.

In accordance with the general formula \eqref{expansion}, the contribution to the $\Omega$-potential from the cut is written as
\begin{multline}\label{omega_cut_msls}
    \pm\beta\Omega^c_d=\sqrt{g}\sideset{}{'}\sum_q\int_0^\infty dss^{1/2}k_{d/2-1}(q,s^{1/2})\\
    =\frac{\sqrt{g}}{4\pi}\sideset{}{'}\sum_q\frac{\Ga\left((d+1)/2\right)}{(\pi q^2)^{(d+1)/2}}=:\frac{\sqrt{g}}{4\pi}\pi^{-(d+1)/2}\Ga\left((d+1)/2\right)\zeta^d_{\mathrm{E}}(g,d+1),
\end{multline}
where $\zeta^d_{\mathrm{E}}(g,\nu)$ is the homogenous Epstein $\zeta$-function. The series defining this function converges at those values of $\nu$ and $d$ which we need. However, it is a slowly convergent series. One can resum this series to write it in the rapidly convergent form. This can be done, for instance, by means of the nice formula from \cite{HJKS,AmbWolf,LimTeo}, but here we apply the method developed in Sec. \ref{Contr_Pole_Real_Axis}. If the eigenvalues of the metric constitute the hierarchy \eqref{eigen_hierarchy} then, in the notation of Sec. \ref{Contr_Pole_Real_Axis}, the resummed homogenous Epstein $\zeta$-function becomes
\begin{multline}\label{zeta_Eps_ex}
    \Ga(\nu)\sideset{}{'}\sum_pp^{-2\nu}\approx\sum_{k=1}^{n+1}\pi^{(\bar{d}-R_k)/2}\sqrt{g_{\bar{d}-R_k}}\biggl\{\frac{\la_k^{\nu-(\bar{d}-R_k)/2}-\la_{k-1}^{\nu-(\bar{d}-R_k)/2}}{\nu-(\bar{d}-R_k)/2}\\
    +\sideset{}{'}\sum_{p_a}p_a^{\bar{d}-R_k-2\nu}\left[\Ga(\nu-(\bar{d}-R_k)/2,\la_{k-1}p_a^2)-\Ga(\nu-(\bar{d}-R_k)/2,\la_kp_a^2)\right]\\
    +\sideset{}{'}\sum_{q_A}(\pi q_A)^{2\nu-\bar{d}+R_k}\left[\Ga((\bar{d}-R_k)/2-\nu,\pi^2 q_A^2/\la_k)-\Ga((\bar{d}-R_k)/2-\nu,\pi^2 q_A^2/\la_{k-1})\right]\biggr\},
\end{multline}
where $\la_k$ are arbitrary real numbers of the order of the eigenvalues $\bar{\la}_k$, and we put by definition $\la_0^\al=\la_{n+1}^\al:=0$, $\forall\al$, and $g_0:=1$, $R_{n+1}=\bar{d}$. An approximate equality means that we neglect strongly exponentially suppressed terms. In the massless case, the ``intermediate'' integrals \eqref{c_k} fall into the third case of Eq. \eqref{c_k_tilde}. Recalling \eqref{eigen_hierarchy}, it is not difficult to see that the argument of the Macdonald function is large and therefore it gives a neglible contribution. It is those terms which we casted out in \eqref{zeta_Eps_ex}. The contributions in the second and third lines of Eq. \eqref{zeta_Eps_ex} are exponentially suppressed. Of course, in view of \eqref{eigen_hierarchy}, the expansion \eqref{zeta_Eps_ex} admits a further simplification, but we leave it intact.

In considering massless particles at the vanishing chemical potential, we restrict ourself to the case of an approximately isotropic metric $g_{ij}$ (all its eigenvalues are of the same order). Therefore, we have the exact expression
\begin{equation}\label{zeta_Eps}
    \sideset{}{'}\sum_q q^{-\nu}=\sideset{}{'}\sum_q\frac{\Ga(\nu/2,\la q^2)}{\Gamma(\nu/2)q^{\nu}}+\frac{\pi^{\nu-d/2}}{\sqrt{g}} \sideset{}{'}\sum_p\frac{\Ga\left((d-\nu)/2,\pi^2p^2/\la\right)}{\Ga(\nu/2)p^{d-\nu}}-\frac{\la^{\nu/2}}{\Ga(\nu/2+1)}+\frac{\pi^{d/2}}{\sqrt{g}}\frac{2\la^{(\nu-d)/2}}{(\nu-d)\Ga(\nu/2)},
\end{equation}
where $\la$ is an arbitrary parameter. It is useful to take its value be of the order of $q_m^{-2}$. Then, making a neglible error, the sums in the RHS of \eqref{zeta_Eps} can be broken off retaining the first several terms only. Notice also that if $L_x$ and $L_y$ are much larger than  $L_z$ and $l_T$ then the series in $q^i$ entering \eqref{omega_cut_msls} becomes effectively one-dimensional over the vectors $q^i$ directed along the $z$-axis and is expressed through the Riemann $\zeta$-function. In this case, the average energy following from \eqref{omega_cut_msls} coincides with the well-known result for the Casimir energy of a one scalar degree of freedom: we must also divide the expression \eqref{omega_cut_msls} on $2^d$ to take into account the Dirichlet boundary conditions.

An expression for the quasiclassical contribution to the $\Omega$-potential can be easily deduced from its integral form. Nevertheless, we shall derive it from the representation \eqref{contrib_q0} as a sum over poles in order to demonstrate the method. The contribution from the cut in  formula \eqref{contrib_q0} is zero for a massless relativistic dispersion law. An analog of the Casimir term in Eq. \eqref{contrib_q0} is proportional to $m^{d+1}$, on restoring the mass, and tends to zero. The second contribution from the cut is understood in the sense of principal value. It is an odd function of $\mu$ regular in a vicinity of zero. Consequently, it vanishes at $\mu\rightarrow0$ too. In the fermionic case, the main (quasiclassical) contribution to the logarithm of the partition function looks like
\begin{equation}
    \beta\Omega^0_d=\frac{\sqrt{g}}{2\pi}(1-2^{-d})(4\pi)^{(d+1)/2}\Gamma\left((d+1)/2\right)\zeta(d+1),
\end{equation}
where we have used the basic functional equation for the $\zeta$-function.

To sum over the poles in the case of the Bose-Einstein distribution, it is necessary to isolate the contribution from the pole $k=0$ since its contribution diverges in the limit $\mu\rightarrow0$. This divergence is expected as long as we sum over $p_i$ in the infinite limits and the contribution to the $\Omega$-potential from $p_i=0$ diverges at a vanishing chemical potential. According to formulas \eqref{contrib_q0_one_pole}, \eqref{contrib_q_one_pole}, and \eqref{contrib_p_one_pole}, this pole contributes to the $\Omega$-potential as
\begin{equation}\label{contrib_one_pole_c_bos}
    2c(d)\approx\pi^{d/2}\sqrt{g}\biggl[\sideset{}{'}\sum_q(\pi q)^{-d}\Ga(d/2,\pi^2q^2/|\la|)-\frac{2}{d|\la|^{d/2}}\biggr]+\sum_p\Ga(0,|\la|(p^2-s_0)),
\end{equation}
where $d$ is the physical space dimension, and we neglect all the terms vanishing at $s_0\rightarrow0$. The contribution from the pole $k=0$ is ``almost'' conformal invariant, i.e., apart from the logarithmic term appearing in the expansion of the incomplete gamma function in a neighbourhood of zero, its contribution is invariant under a simultaneous dilatation of all the metric components $g_{ij}$. A conformal invariant part of the $\Omega$-potential does not contribute to the average energy.

It is useful to assign the parameter $\la$ to be equal to $-iq^2_m$. This choice allows us to terminate both the sum over $q$ and the sum over $p$ keeping a few leading terms from these series. A contribution from the rest of poles is found in the same way as in the fermionic case considered above. The oscillating correction to the $\Omega$-potential without the zeroth pole contribution can be cast into the form
\begin{equation}
    \beta\Omega^{os}_d\approx\frac{\sqrt{g}}{2\pi}\sideset{}{'}\sum_q\left(\frac{2\pi}{q}\right)^{(d+1)/2}\sum_{s=0}^\infty\frac{(8\pi^2q)^{-s}\Ga\left((d+1)/2+s\right)}{s!\Ga\left((d+1)/2-s\right)}\Li_{s-(d-1)/2}\left(e^{-4\pi^2q}\right),
\end{equation}
while the quasiclassical contribution reads as
\begin{equation}
    \beta\Omega^0_d=\frac{\sqrt{g}}{2\pi}(4\pi)^{(d+1)/2}\Gamma\left((d+1)/2\right)\zeta(d+1),
\end{equation}
where we neglect all the terms disappearing in the limit $\mu\rightarrow0$. As a result, in the bosonic case
\begin{equation}
    \beta\Omega_d=\beta\Omega^0_d+\beta\Omega^c_d+\beta\Omega^{os}_d+c(d).
\end{equation}
All the above mentioned conclusions regarding the oscillating contribution to the fermionic $\Omega$-potential hold for bosons as well. We also see that the oscillating contribution in the bosonic case is more suppressed than in the fermionic case. In both cases, it grows exponentially with decreasing sizes of the system and becomes appreciable on the scales of the order of $l_T$. The expression for the main contribution to the $\Omega$-potential both for fermions and bosons coincides with the leading term of the high-temperature expansions \eqref{high_temp_ferm} and \eqref{high_temp_bos}.

Now we take into account that, at zero boundary conditions on the fields, the sums defining the partition function are not in the infinite limits, but from unity to infinity. Introducing a notation for the sums and recalling a symmetry of the spectrum with respect to the substitution $p_i\rightarrow-p_i$ for any quantum number, we arrive at
\begin{equation}\label{reduction_formul}
    \sum_{p_1=1}^\infty\cdots\sum_{p_d=1}^\infty f(p^2)=2^{-d}\sum_A (-1)^{d-n}\sum_{p_{\al_1}=-\infty}^\infty\cdots\sum_{p_{\al_n}=-\infty}^\infty f(p^2),
\end{equation}
where $(\al_1,\ldots,\al_n)$, $\al_i=\overline{1,d}$, is a collection of nonequal natural numbers in the increasing order. The summation is carried over all such collections including the empty one. There are $2^d$ such collections in $d$ dimensions. An absence of the sum over some $p_i$ in the RHS of \eqref{reduction_formul} means that $p_i=0$.

Consider separately the zero dimensional contribution entering the RHS of Eq. \eqref{reduction_formul}. In the case where all $p_i=0$, we have for bosons
\begin{equation}\label{zero_dimen_exact}
    \beta\Omega_0=-\ln(1-e^\mu).
\end{equation}
At $\mu\rightarrow0$, the divergent contributions to the bosonic $\Omega$-potential are the term \eqref{zero_dimen_exact} and the contributions from the second term in Eq. \eqref{contrib_one_pole_c_bos} taken in different dimensions excepting zeroth one\footnote{The zeroth dimension is excluded since, in this dimension, the whole $\Omega$-potential is given by formula \eqref{zero_dimen_exact}}. It is not difficult to comprehend that a passage to the semi-infinite summation intervals reduces for these last contributions to the change of summation limits which also become semi-infinite, i.e., from unity to infinity. However, to this end, it is necessary to add to the sum the term of such a type taken in zeroth dimension,
\begin{equation}
  2^{-1-d}(-1)^d\Ga(0,-q^2_ms_0),
\end{equation}
because it is not contained in \eqref{reduction_formul}, and subtract it. Collecting this subtrahend term with the zeroth dimension contribution \eqref{zero_dimen_exact}, we obtain
\begin{equation}
  2^{-d}(-1)^{d+1}(\Ga(0,-q^2_ms_0)/2+\ln(1-e^\mu))\underset{\mu\rightarrow0}{\rightarrow}\frac{(-1)^d}{2^{d+1}}(\ga_{\mathrm{E}}+\ln q^2_m).
\end{equation}
Then, denoting by $A_n:=(\al_1,\ldots,\al_n)$, by $g^{A_n}_{ij}$ the respective block ($n\times n$) of the initial metric and so on, we come to the rapidly convergent expansion for the bosonic $\Omega$-potential
\begin{multline}\label{omega_msls_bos}
    \beta\Omega_d=
    \sum_A(-1)^{d-n}\sqrt{g_{A_n}}\biggl[\frac{\pi^{(n-1)/2}}{2^{d-n}}\Gamma\left((n+1)/2\right)\zeta(n+1)-\frac{2^{-d}\pi^{n/2}}{nq_m^n}+\sideset{}{'}\sum_{q_{A_n}}\frac{\Ga(n/2,\pi^2q^2_{A_n}/q^2_m)}{2^{d+1}(\pi q^2_{A_n})^{n/2}}\\
    +\sideset{}{'}\sum_{q_{A_n}}\left(\frac{2\pi}{q_{A_n}}\right)^{(n+1)/2}\sum_{s=0}^\infty\frac{(8\pi^2q_{A_n})^{-s}\Ga\left((n+1)/2+s\right)}{2^{d+1}\pi s!\Ga\left((n+1)/2-s\right)}\Li_{s-(n-1)/2}\left(e^{-4\pi^2q_{A_n}}\right)\biggr]\\
    +\frac{(-1)^d}{2^{d+1}}(\ga_{\mathrm{E}}+\ln q^2_m)
    +\sum_{p_1\cdots p_d=1}^\infty\frac{\Ga(0,q^2_mp^2)}2-\sum_A(-1)^{d-n}\sqrt{g_{A_n}}\frac{\Ga((n+1)/2)}{2^{d+2}\pi^{(n+3)/2}}\zeta^n_{\mathrm{E}}(g_{A_n},n+1),
\end{multline}
where the summation over collections $A$ no longer includes the empty one.

\begin{figure}[t]
\centering
\includegraphics*[trim=0mm 15mm 0mm 15mm,clip,width=7cm]{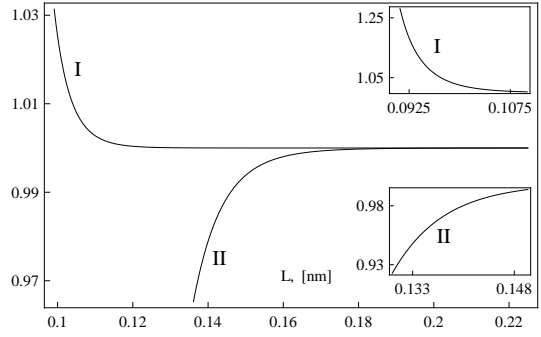}%{m-less-3.eps}
\caption{{\footnotesize The ratio of the exact value of the $\Omega$-potential taken from its definition to its approximate value given by Eq. \eqref{omega_msls_bos} for bosons (I) and Eq. \eqref{omega_msls_fer} for fermions (II). As expected, the asymptotic expansions \eqref{omega_msls_bos} and \eqref{omega_msls_fer} deviate from the exact values when the condition \eqref{cond rel_high} is violated.}}\label{plot massless}
\end{figure}

As far as the massless fermions are concerned there are not any problems with divergencies of separate terms at zero chemical potential like in the bosonic case. So, we immediately write for the $\Omega$-potential
\begin{multline}\label{omega_msls_fer}
    \beta\Omega_d=
    (-2)^{-d}\ln2+\sum_A(-1)^{d-n}\sqrt{g_{A_n}}\biggl[\frac{\pi^{(n-1)/2}}{2^{d-n}}(1-2^{-n})\Gamma\left((n+1)/2\right)\zeta(n+1)\\
    +\frac{\Ga((n+1)/2)}{2^{d+2}\pi^{(n+3)/2}}\zeta^n_{\mathrm{E}}(g_{A_n},n+1)
    +\sideset{}{'}\sum_{q_{A_n}}\left(\frac{2\pi}{q_{A_n}}\right)^{(n+1)/2}\sum_{s=0}^\infty\frac{(8\pi^2q_{A_n})^{-s}\Ga\left((n+1)/2+s\right)}{2^{d+1}\pi s!\Ga\left((n+1)/2-s\right)}\\
    \times\Bigl[\Li_{s-(n-1)/2}\left(e^{-4\pi^2q_{A_n}}\right)-2^{s-(n-1)/2}\Li_{s-(n-1)/2}\left(e^{-2\pi^2q_{A_n}}\right)\Bigr]\biggr],
\end{multline}
where we explicitly single out the contribution from zeroth dimension. Remind that, if the above mentioned conditions are fulfilled, all the sums in the obtained expansions can be broken off retaining several leading terms.  Besides, we can eliminate a summation of identical terms in the sums over $q$ in formulas \eqref{omega_msls_bos} and \eqref{omega_msls_fer}. It is easy to prove the following combinatorial relation
\begin{equation}
    2^{-d}\sum_A (-1)^{d-n}\sideset{}{'}\sum_{q_{A_n}=-\infty}^\infty f_n(q^2_{A_n})=\sum_A\sum_{q_{A_n}=1}^\infty\sum_{k=0}^{d-n}2^{n-d}(-1)^kC_{d-n}^k f_{d-k}(q^2_{A_n}),
\end{equation}
where $f_n(x)$ are arbitrary functions. Here we have exploited a metric diagonality. The contribution from the empty collection is absent both in the left and RHS of the equality. In particular, if $f_n(x)$ is independent of $n$ then the RHS reduces to the $d$-dimensional sum in semi-infinite limits.

A comparison of the obtained asymptotic expansions for the $\Omega$-potential with its exact value is presented on Fig. \ref{plot massless}.

\subsubsection{Nonvanishing chemical potential}

Now we briefly describe how the above results change at the non-zero (electro)chemical potential. We shall consider only the fermionic case bearing in mind the electrons in graphene (graphite). Near the Dirac points, the dispersion law of the electrons and holes looks like \cite{KatNov,NGMJKGDF,NGPNG}
\begin{equation}\label{disper_law_gr}
    \e(\spp)=\tilde{\mu}\pm\ups_F|\spp|,
\end{equation}
where $\tilde{\mu}\approx0.3$ eV, and $\ups_F\approx9.1\times10^7$ cm/s. We interpret the first term in the dispersion law \eqref{disper_law_gr} as the chemical potential. This interpretation is correct only if the form of the dispersion law does not change during the studied process (a deformation, in our case). Although we shall assume a constancy of the first term in \eqref{disper_law_gr}, corrections for the variations of this term can be directly included to the final expressions for the $\Omega$-potential and average charge thereby relaxing this assumption. Besides, it was shown in \cite{SoCoLo,GuLiZho} that the dispersion law \eqref{disper_law_gr} has a gap which can be modeled by a small mass entering the dispersion law. This gap also depends on whether the graphene lattice is deformed or not. We completely neglect this contribution to the dispersion law as long as we are going to consider a graphene specimen with the armchair edges where this gap is absent in the tight-binding approximation. Thus, at the temperature $\beta^{-1}=1$ K, we have the dimensionless chemical potential $\mu\approx3.5\times 10^3$ and the thermal wave length $l_T\approx2.2\times 10^{-3}$ cm. In graphene, there are two branches of the dispersion law \eqref{disper_law_gr} and also two spin degrees of freedom for each of the branches, i.e., in sum, eight degrees of freedom. For simplicity, we shall study a ribbon specimen with sizes $(L_x,L_y)$, a length of the ribbon $L_y$ being large: $L_y\gg l_T$. Then, for the armchair boundary conditions \cite{BreFer,NakFuj,NGPNG}
\begin{equation}\label{quant_cond_gr}
    p_x=\frac{\pi\hbar}{L_x}(n+4k/3),\qquad n\in\Z,\quad k=\{0,1,2\},
\end{equation}
where $k=L_x/a_0\mod3$ and $a_0\approx1.42\times10^{-8}$ cm is a length of the lattice translation vector. The eigenfunctions of the one-particle Hamiltonian implying this quantization condition are represented by a composition of two wave functions corresponding to the different Dirac points. Therefore, the degeneracy over the spin degrees of freedom only survives.

The quasiclassical contribution to the $\Omega$-potential is expressed through the Hurwitz $\zeta$-function (see, e.g., \cite{BatErde}) and can be cast into the form
\begin{multline}\label{omega_qc_gr}
    \beta\Omega_d^0=\frac{4\sqrt{g}}{\pi}(4\pi)^{(d+1)/2}\Ga\left((d+1)/2\right)\biggl\{\sum_{k=0}^{[(d+1)/2]}\frac{\zeta(2k)(1-2^{1-2k})}{\Ga(d+2-2k)}\mu^{d+1-2k}\\
    +\frac{1+(-1)^d}{2}\left[\Li_{d+1}\left(e^{-\mu}\right)-2^{-d}\Li_{d+1}\left(e^{-2\mu}\right)\right] \biggr\},
\end{multline}
where the spin degrees of freedom are taken into account. The contribution from the cut $\Omega_d^c$ is independent of the chemical potential and has the form \eqref{omega_cut_msls} multiplied by
\begin{equation}
    8\cos(2\pi k/3).
\end{equation}
The oscillating term is given by
\begin{equation}\label{omega_os_gr}
    \beta\Omega^{os}_d\approx-4\pi i\sqrt{g}\sideset{}{'}\sum_{q^1}e^{2\pi ikq^1/3}\sum_{k=0}^\infty [h^{(1)}_{d/2}(q_m |q^1|,i\omega_k+\mu)+h^{(1)}_{d/2}(q_m|q^1|,i\omega_k-\mu)],
\end{equation}
where $q_m=L_x/l_T$ and the exponentially suppressed contributions from macroscopic dimensions are discarded. Recall that only the poles corresponding to the electron branch of the dispersion law \eqref{disper_law_gr} contribute to the oscillating part. The holes' poles lie on the ``unphysical'' sheet. We simplify this expression under the assumption that the condition \eqref{Hankel_func_arg} is fulfilled for the zeroth pole:
\begin{equation}
    4\pi q_m|\mu+i\pi|\gg1.
\end{equation}
At the value of the chemical potential presented above and the temperature $\beta^{-1}=1$ K, this condition implies that we are about to consider the ribbon specimen with a width larger than $5\times10^{-7}$ cm. Then, making use of the asymptotic expansion of formula \eqref{onedim_sum}, we come to
\begin{equation}\label{omega_os_gr_1}
    \beta\Omega^{os}_d\approx-\frac{16\sqrt{g}}{q_m^{(d+1)/2}}\sum_{n=0}^\infty\frac{\Ga((d+1)/2+n)}{n!\Ga((d+1)/2-n)}\sum_{k=0}^\infty\re\left[\frac{(\omega_k-i\mu)^{(d-1)/2}}{(4\pi q_m(\omega_k-i\mu))^n}\Li_{(d+1)/2+n}\left(e^{-2\pi q_m(\omega_k-i\bar{\mu})}\right)\right],
\end{equation}
where $\bar{\mu}:=\mu+k/3q_m$. The respective contributions to the conserved charge $Q$ are easily derived from expressions \eqref{omega_qc_gr} and \eqref{omega_os_gr} with the help of the recurrence relation \eqref{recurr relat}. As a result,
\begin{multline}\label{Q_qc_gr}
    Q_d^0=\frac{4\sqrt{g}}{\pi}(4\pi)^{(d+1)/2}\Ga\left((d+1)/2\right)\biggl\{\sum_{k=0}^{[d/2]}\frac{\zeta(2k)(1-2^{1-2k})}{\Ga(d+1-2k)}\mu^{d-2k}\\
    -\frac{1+(-1)^d}{2}\left[\Li_d\left(e^{-\mu}\right)-2^{1-d}\Li_d\left(e^{-2\mu}\right)\right] \biggr\},
\end{multline}
and
\begin{equation}\label{Q_os_gr}
    Q^{os}_d\approx\frac{32\pi\sqrt{g}}{q_m^{(d-1)/2}}\sum_{n=0}^\infty\frac{\Ga((d-1)/2+n)}{n!\Ga((d-1)/2-n)}\sum_{k=0}^\infty\im\left[\frac{(\omega_k-i\mu)^{(d-1)/2}}{(4\pi q_m(\omega_k-i\mu))^n}\Li_{(d-1)/2+n}\left(e^{-2\pi q_m(\omega_k-i\bar{\mu})}\right)\right].
\end{equation}
These equations implicitly define the chemical potential $\tilde{\mu}$ at the fixed value of the charge $Q$. Under normal conditions (unperturbed crystal), this charge is determined by the above mentioned value of the chemical potential $\tilde{\mu}\approx0.3$ eV. Of course, there is not any uncompensated charge in the crystal under the normal conditions. The non-zero charge $Q$ results from the chemical potential redefinition made by us.

The expansions \eqref{omega_os_gr} and \eqref{Q_os_gr} are slowly convergent with respect to $k$ when
\begin{equation}\label{smallness_cond}
    2\pi^2 q_m\ll1.
\end{equation}
For example, at the temperature $\beta^{-1}=1$ K and the ribbon width $L_x=10^{-6}$ cm, the value of this parameter becomes $2\pi^2q_m\approx9.1\times10^{-3}$. If the condition \eqref{smallness_cond} takes pace, we may apply the Euler-Maclaurin formula to Eq. \eqref{Q_os_gr} to arrive at
\begin{equation}\label{Q_os_gr_1}
    Q^{os}_d\approx\frac{8\sqrt{g}}{\pi q_m^{(d+1)/2}}\sum_{n=0}^\infty\frac{\Ga((d+1)/2+n)}{n!\Ga((d+1)/2-n)}\im\left[\frac{(\pi-i\mu)^{(d-1)/2}}{(4\pi q_m(\pi-i\mu))^n}\Li_{(d+1)/2+n}\left(e^{-2\pi q_m(\pi-i\bar{\mu})}\right)\right].
\end{equation}
In order to take properly into account the boundary conditions chosen by us, we must divide the expressions for the $\Omega$-potential and the charge $Q$ on $2^d=4$. Hence, neglecting the exponentially suppressed terms in \eqref{Q_qc_gr}, we finally have
\begin{equation}
\begin{split}
    Q^0&\approx\pi\frac{L_xL_y}{l_T^2}(\mu^2+\pi^2/3),\\
    Q^{os}&\approx\frac{2L_y}{\pi l_T q_m^{1/2}}\sum_{n=0}^\infty\frac{\Ga(3/2+n)}{n!\Ga(3/2-n)}\im\left[\frac{(\pi-i\mu)^{1/2}}{(4\pi q_m(\pi-i\mu))^n}\Li_{3/2+n}\left(e^{-2\pi q_m(\pi-i\bar{\mu})}\right)\right].
\end{split}
\end{equation}
There are not the contributions from lower dimensions in the case at hand. Note that the obtained formulas hold in the zero-temperature limit $\beta\rightarrow\infty$ and allow simply to find the zero-temperature asymptotics of the charge $Q$.

\begin{figure}[t]
\centering
\includegraphics*[trim=0mm 15mm 0mm 15mm,clip,width=7cm]{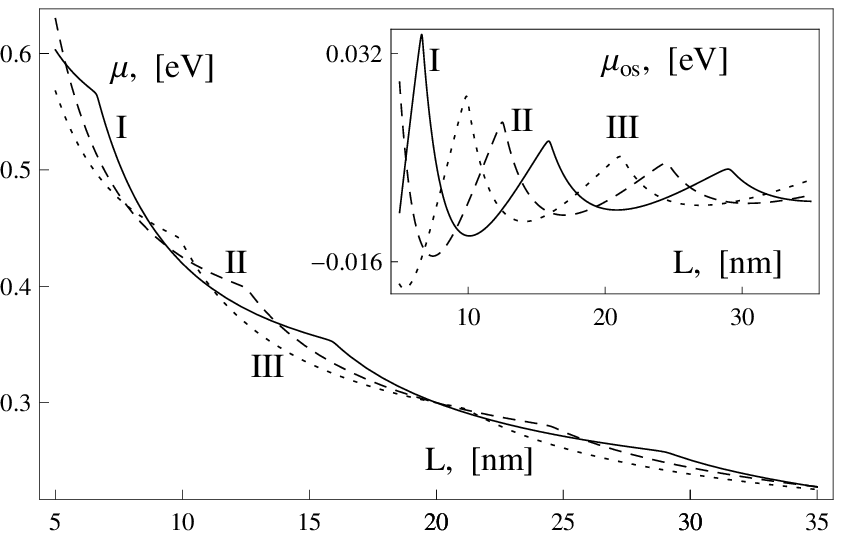}\;%{grafen-2.eps}\;
\includegraphics*[trim=0mm 15mm 0mm 15mm,clip,width=7cm]{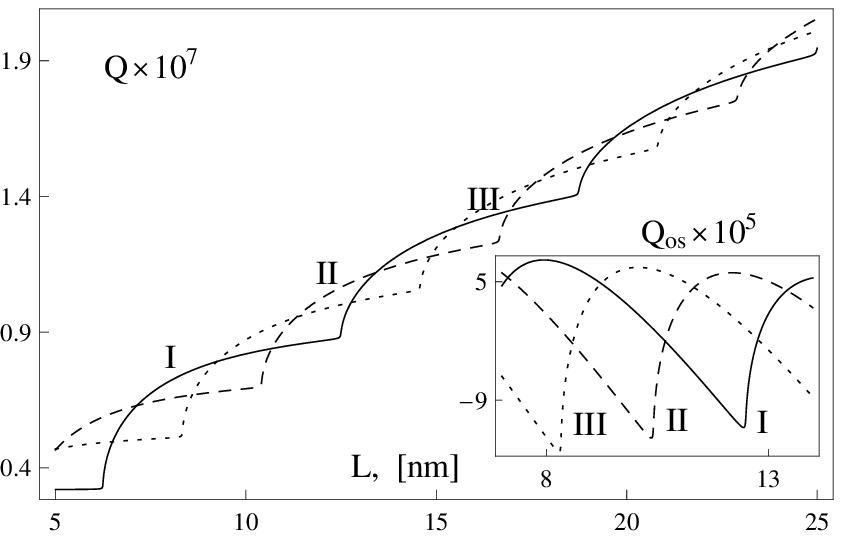}

\caption{On the left panel: The total chemical potential of electrons in the graphene ribbon of the length $L_y=1$ cm at $\beta^{-1}=1$ K. The undeformed crystal (normal conditions) is of the width $L_x=20$ nm and at the chemical potential $\tilde{\mu}=0.3$ eV. The parameter $k$ entering Eq. \eqref{quant_cond_gr} is set to $k=0$ (I), $k=1$ (II), and $k=2$ (III). The small plots depict the corresponding oscillating parts of the chemical potential. On the right panel: The total charge $Q$ of electrons in the graphene ribbon of the length $L_y=1$ cm at the fixed chemical potential $\tilde{\mu}=0.3$ eV, the temperature $\beta^{-1}=1$ K, and $k=0$ (I), $k=1$ (II), and $k=2$ (III). The small plots depict the corresponding oscillating parts of the charge.}
\label{plot graphene}
\end{figure}

Now it is not difficult to describe the basic features of oscillations of the chemical potential. Expressing perturbatively $\mu$ from the above equations, we find that the period of oscillations of the dimensionless chemical potential with $L_x$ reads as
\begin{equation}
    \ell\approx2\sqrt{\frac{\pi L_xL_y}{Q}}.
\end{equation}
It grows with $L_x$  starting with the value $2l_T/\mu_0$, where $\mu_0$ is the dimensionless chemical potential at the normal conditions. This value is twice greater in comparison with the naive estimation. Keeping in mind the condition \eqref{smallness_cond}, we obtain that the amplitude of these oscillations is of the order of
\begin{equation}
    |\mu_{os}|\approx\zeta(3/2)\left(\frac{L_y}{4\pi^7Ql_Tq_m^5}\right)^{1/4}.
\end{equation}
It decreases with $L_x$ starting with the value
\begin{equation}
    \zeta(3/2)\left(2\pi^4q_m^3\mu_0\right)^{-1/2},
\end{equation}
where $q_m$ corresponds to the unperturbed crystal. We see from these formulas that the order of the amplitude and period of oscillations of the chemical potential $\tilde{\mu}$ does not appreciably depend on the temperature. The plots of these oscillations are presented on Fig. \ref{plot graphene}. The oscillation period is rather large at the value of the chemical potential $\tilde{\mu}$ given above. So, the oscillations of this type cannot be seemingly observed as oscillations and not as the one peak. Inasmuch as the period of these oscillations decreases with the width of a ribbon, they are likely to be observed under a squeezing of the specimen.

\subsection{Electrons in a thin metal film}\label{Ele_Th_Me_Fi}

In this section, we consider thermodynamic properties of the conduction electrons in a thin metal film. This is a classical problem for the Fermi-liquid theory of metals. It was studied in detail both for the quadratic dispersion law (refs \cite{Pieirls,BregZhuch,Rumer,Zilb}) and, in the quasiclassical approximation, for the dispersion law of an arbitrary form (see, e.g., \cite{OnsagOsc,Shoenb,Nedore,Azbel,Kulik,LifshKag}). It is well-known that the thermodynamic properties of such electrons are described by the Fermi-Dirac distribution with a satisfactory accuracy. The dimensionless chemical potential $\mu$ entering the distribution should be expressed through the number of electrons in the conduction band under the assumption that this number is conserved. We shall investigate the metals with the only one conduction band, but all the below results are straightforwardly generalized to the case of several conduction bands. Also we restrict ourself to the isotropic (in terms of the quasi-momentum) dispersion law and derive explicit rather simple expressions for the $\Omega$-potential and the number of conduction electrons which are valid with an exponential accuracy. This approximation is adequate for the electrons located in a sufficiently small vicinity of the conduction band bottom. An analogous but less elaborated analysis for a cubic specimen of a metal is given in \cite{IzvVuz}.

Before we proceed to a calculation of the partition function, we have to make some remarks on the way how we are going to take into account a finiteness of the crystal. A presence of a surface on the crystal changes appreciably the properties of this crystal near the surface, deforms its lattice, and even can lead to a lowering of the two-dimensional symmetry of the bulk specimen. A description of these effects from the first principles is quite difficult and, in the most cases, does not admit an analytic investigation. Hence, we shall suppose that the defects of the crystal lattice appearing due to a presence of the surface do not influence the thermodynamical properties of the conduction electrons too much. We shall consider the simplest model of a finite crystal.

Let $V(\mathbf{x})$ be the effective potential of the electron for the infinite periodic lattice
\begin{equation}
    V(\mathbf{x}+n^i\mathbf{a}_i)=V(\mathbf{x}),\qquad n^i\in\Z,
\end{equation}
where $\mathbf{a}_i$ are the lattice translation vectors. A complete set of solutions to the Schr\"{o}dinger equation for the electron is constituted by the Bloch wave functions
\begin{equation}\label{Schrod_eq}
    \left(-\frac{\hbar^2}{2m}\Delta+V(\mathbf{x})\right)\psi_{s\mathbf{p}}(\mathbf{x})=\e_s(\mathbf{p})\psi_{s\mathbf{p}}(\mathbf{x}),
\end{equation}
where $m$ is the electron mass, $s$ is the band index, $\mathbf{p}$ is the quasi-momentum, and $\e_s(\mathbf{p})$ is the dispersion law. The Bloch functions are defined in the standard way
\begin{equation}
    \psi_{s\mathbf{p}}(\mathbf{x})=e^{i\mathbf{p}\mathbf{x}/\hbar}u_{s\mathbf{p}}(\spx),\qquad\psi_{s\mathbf{p}+2\pi\hbar\mathbf{b}^i}(\spx)=\psi_{s\mathbf{p}}(\spx),\qquad u_{s\mathbf{p}}(\spx+\mathbf{a}_i)=u_{s\mathbf{p}}(\spx),
\end{equation}
where $\mathbf{b}^i$ are the basis vectors of the reciprocal lattice: $\mathbf{a}_i\mathbf{b}^j=\de_i^j$. For brevity, we neglect the spin properties of the electron and shall take them into account only in counting the degeneracy order of energy levels. Now we cut out from the infinite crystal a parallelepiped spanned on the basis vectors $\mathbf{a}_i$ multiplied by some natural numbers. We shall assume that the effective potential of the electron in this finite crystal differs negligibly from $V(\mathbf{x})$.

In that case, the solutions to the Schr\"{o}dinger equation obeying the Dirichlet boundary conditions can be obtained by a linear combination of the Bloch waves $\psi_{s\spp}$ corresponding to the same energy $\e_s(\spp)$. It is useful to choose the basis $\{\mathbf{a}_i\}$ in the $\spx$-space and the basis $\{\mathbf{b}^i\}$ in the reciprocal $\spp$-space. Then this boundary conditions imply the following system of equations on the coefficients of the linear combination $c_{s\spp}$:
\begin{equation}\label{boundar_cond}
    \sum_{s p_i}c_{s\spp}\left.u_{s\spp}(\spx)\right|_{x^i=0}=0,\qquad \sum_{s p_i}e^{in_ip_i/\hbar}c_{s\spp}\left.u_{s\spp}(\spx)\right|_{x^i=0}=0,\qquad i=\overline{1,3},
\end{equation}
where the vectors $\spp$ lie on the constant-energy surface $\e_s(\spp)=\e_s$, while $n_i$ are natural numbers specifying a number of unit cells along each of the basis vectors $\mathbf{a}_i$ in the finite crystal. Suppose that the equation,
\begin{equation}\label{isoenerg_surf}
    \e_s(\spp)=\e_s,
\end{equation}
on the quasi-momentum component in the chosen basis $p_i$ has two or less solutions $p^{(1,2)}_i(\e_s,\spp_{\perp})$ for every $i$. Together with Eqs. \eqref{boundar_cond}, this requirement results in the quantization condition \cite{Nedore,LifshKag}
\begin{equation}\label{quant_cond}
    |p^{(2)}_i-p^{(1)}_i|=2\pi\hbar\frac{k_i}{n_i},\qquad k_i\in\mathbb{N}.
\end{equation}
In order to satisfy this condition, one need to plot in the $\spp$-space a parallelepiped with edges directed along the basis vectors $\mathbf{b}^i$. The lengths of its edges must be equal to \eqref{quant_cond} multiplied by the length of the corresponding basis vector $\mathbf{b}^i$ and all its vertices should lie on the constant-energy surface \eqref{isoenerg_surf}. Of course, it is not possible for any value $\e_s$. If the form of the cut crystal does not agree with its periodic structure, the conditions \eqref{quant_cond} ought to be understood as a quasiclassical approximation to the exact quantization conditions.

As we have already mentioned, we shall consider the electrons with quadratic and isotropic dispersion law in the single conduction band
\begin{equation}
    \e(\spp)=\frac{\spp^2}{2m_*},
\end{equation}
where $m_*$ is an effective electron mass. Also we assume that the reciprocal lattice is cubic. Nevertheless, all the below formulas are immediately generalized by a redefinition of the constants entering them to the case where the constant-energy surfaces of the dispersion law \eqref{isoenerg_surf} have the forms of ellipsoid with the principal axes directed along the reciprocal basis vectors. For the given dispersion law, the quantization conditions \eqref{quant_cond} look in the Cartesian basis like
\begin{equation}
    p_i=\frac{\pi\hbar k_i}{L_i},\qquad k_i\in\mathbb{N},
\end{equation}
where $L_i$ are sizes of the crystal. Passing to the notation used in \eqref{part func}, we see that, in our case, the metric takes the form
\begin{equation}
    g_{ij}=l_D^{-2}diag(L_x^2,L_y^2,L_z^2),\qquad l_D:=\left(\frac{\beta\pi^2\hbar^2}{2m_*}\right)^{1/2},
\end{equation}
where $l_D$ is the thermal de Broglie wavelength. At the temperature $\beta^{-1}=1$ K and the effective electron mass $m_*$ equal to its mass in vacuum, the thermal wavelength is approximately $6.6\times10^{-6}$ cm. The roots of Eq. \eqref{zeros} become (the parameter $m$ in Eq. \eqref{zeros} is put to zero)
\begin{equation}\label{singularities_expl}
    p_k=\frac{1}{\sqrt{2}}\left(\sgn(k)\sqrt{\sqrt{\mu^2+\pi^2(2k+1)^2}+\mu}+i\sqrt{\sqrt{\mu^2+\pi^2(2k+1)^2}-\mu}\right),\quad k\in\Z,
\end{equation}
where the sign function in zero is defined as $\sgn(0)=1$. In accordance with our notation, the reciprocal temperature $\beta$ is included to the dimensionless chemical potential $\mu$. At the temperature $\beta^{-1}=1$ K and the chemical potential $\tilde{\mu}=11.7$ eV (Aluminium), we have for the dimensionless chemical potential $\mu\approx1.36\times10^5$.

First, consider the case where a summation over the quantum number in the logarithm of the partition function is carried over the infinite limits $(-\infty,\infty)$ . In formula \eqref{ID0-true}, we derived the quasiclassical contribution for this case
\begin{equation}
    \beta\Omega^0_d=-\pi^{d/2}\sqrt{g}\Li_{d/2+1}(-e^\mu).
\end{equation}
The oscillating contribution is given by the general formula \eqref{expansion}. For the nonrelativistic dispersion law, the distance between poles $|p_{k+1}-p_k|$ decreases with an increase of the absolute value of the number $k$, and $\im p_k$ grows. Furthermore, the greater the value of $\mu$, the lesser the distance between adjacent poles. That is why we can sum over the poles making use of the Euler-Maclaurin formula \eqref{Idq_Eu_Mac}. In the case at hand, we have two subsequences $s_\al$ that may be summed. These are the poles $s_k$ with negative and nonpositive $k$, respectively. Thereby, we come to
\begin{equation}\label{Id_nonrel_Hank}
    I^q_d\approx2\pi\sqrt{g}\im\left[\sum_{k=0}^{K-1}h^{(1)}_{d/2}(q,p_k)+\frac{i}{2\pi^2}h^{(1)}_{d/2+1}(q,p_K)\right]+\ldots
\end{equation}
To find the oscillating part of the $\Omega$-potential, we need to sum this expression over $q$. Now we use the fact that we consider a thin metal film, viz., we suppose that $L_y$ and $L_z$ are much larger than $L_x$ and $l_D$.  This assumption considerably simplifies the subsequent formulas for the $\Omega$-potential and the average number of electrons. In virtue of this assumption, the terms of the sum over $q$ with the vectors $q^i$ possessing nonvanishing components along $y$ and $z$ axes are strongly exponentially suppressed (see the expansion of the Hankel function \eqref{Hankel_func_expans}). Then the sum over $q$ reduces effectively to a one-dimensional one.

In principle, we can apply to the sums over $q$ of the terms in Eq. \eqref{Id_nonrel_Hank} the method developed in Sec. \ref{Contr_Pole_Real_Axis}, but we shall act in other way. We shall derive a simpler looking expansion under the assumption that the condition \eqref{Hankel_func_arg} is fulfilled for $q=q_m$ and $p=p_0$. For example, at the temperature $\beta^{-1}=1$ K, the chemical potential $\tilde{\mu}=11.7$ eV, and the film thickness $L_x=10^{-8}$ cm,
\begin{equation}
    4\pi q_m|p_0|\approx7.
\end{equation}
So, this assumptions is quite reasonable. If $q_m$ is so much small that the condition \eqref{Hankel_func_arg} is violated then the rapidly convergent expansion of the $\Omega$-potential can be deduced directly from formula \eqref{part func}. In that case, we just integrate over macroscopic dimensions and cast out the higher terms of the series in the quantum number corresponding to the microscopic dimension.

Thus, supposing the condition \eqref{Hankel_func_arg} is satisfied and using Eq. \eqref{onedim_sum}, we get
\begin{multline}
    \beta\Omega_d^{os}\approx4\pi\sqrt{g}\im \sum_{n=0}^\infty e^{-i\pi(d+1)/4}\left(\frac{i}{4\pi}\right)^n\frac{\Ga\left((d+1)/2+n\right)}{n!\Ga\left((d+1)/2-n\right)}\\
    \times\left[\sum_{k=0}^{K-1}\frac{p_k^{(d-1)/2-n}}{q_m^{(d+1)/2+n}}\Li_{(d+1)/2+n}\left(e^{2\pi iq_mp_k}\right)+\frac{p_K^{(d+1)/2-n}}{2\pi^2q_m^{(d+3)/2+n}}\frac{d+1+2n}{d+1-2n}\Li_{(d+3)/2+n}\left(e^{2\pi iq_mp_K}\right)\right],
\end{multline}
where $q_m=L_x/l_D$. Note that, in odd dimensions, the sum over $n$ terminates and the fulfillment of the condition \eqref{Hankel_func_arg} is not necessary.

Now we should use formula \eqref{reduction_formul} for a passage to the semi-infinite summation limits. As long as $L_y$ and $L_z$ are much greater than $l_D$, the lower dimensions give small contributions to the partition function. For the quasiclassical contribution, we take into account only the first correction due to lower dimensional sums. As for the oscillating term, we retain the leading contribution only. This approximation is justified when the condition \eqref{Hankel_func_arg} is satisfied and
\begin{equation}
    \max(1,\sqrt{\mu})L_y/l_D\gg1,\qquad(|p_0|/q_m)^{1/2}L_y/l_D\gg1.
\end{equation}
The first requirement allows us to discard the contributions from lower dimensions to the quasiclassical part of the $\Omega$-potential, while the second condition permits to make the same with the oscillating terms.

Then, recovering the spin degrees of freedom of the electrons, we arrive at
\begin{equation}
    \beta\Omega^0\approx-\frac{\pi S}{4l_D^2}[\sqrt{\pi}q_m\Li_{5/2}(-e^\mu)-\Li_2(-e^\mu)],
\end{equation}
where $S:=L_yL_z$. The sum over $n$ in the oscillating contribution terminates for $d=3$. So, we obtain
\begin{multline}\label{omega_os}
    \beta\Omega^{os}\approx-\frac{\pi S}{q_ml_D^2}\im\biggl\{\sum_{k=0}^{K-1}\left[p_k\Li_2(e^{2\pi iq_mp_k})+\frac{i}{2\pi q_m}\Li_3(e^{2\pi iq_mp_k})\right]\\
    +\frac{p_K^2}{2\pi^2 q_m}\Li_3(e^{2\pi iq_mp_K})+\frac{3ip_K}{4\pi^3q^2_m}\Li_4(e^{2\pi iq_mp_K})-\frac{3}{8\pi^4q^3_m}\Li_5(e^{2\pi iq_mp_K})\biggr\}.
\end{multline}
The number $K$ entering this formula should be increased until the value of $\Omega^{os}$ ceases to change considerably. At large dimensionless chemical potentials $\mu$, as it takes place for the conduction electrons in a metal, the Euler-Maclaurin formula is valid with a great accuracy and the parameter $K$ can be set to unity or even to zero. The obtained expression is an oscillating function of the size $L_x$. The periods of oscillations at the fixed dimensionless chemical potential are approximately equal to
\begin{equation}\label{frequencies}
    \ell_k\approx l_D/\re p_k,\quad k=0,1,\ldots
\end{equation}
If the imaginary part of the roots $p_k$ grows rapidly with $k$ then the contributions with the large numbers $k$ are strongly exponentially suppressed. In that case, the oscillations mainly occur at the frequencies with $k=0$ and $k=1$. Of course, this also holds at $K=1$. At sufficiently low temperatures, $\mu\gg1$, we obtain a physically expected value for the oscillation period
\begin{equation}
    \ell\approx\frac{\pi\hbar}{\sqrt{2m_*\tilde{\mu}}}.
\end{equation}
At the temperature $\beta^{-1}=1$ K, the chemical potential $\tilde{\mu}=11.7$ eV, and $m_*=m$, the oscillation period becomes $\ell\approx1.8\times 10^{-8}$ cm.

Now we shall find the number of conduction electrons in a metal. To this end, we need to differentiate the obtained expression for the $\Omega$-potential with respect to $\mu$. It is easy to do by use of the recurrence relation
\begin{equation}
    \partial_\mu I_d^q=\pi I^q_{d-2},
\end{equation}
that holds for the nonrelativistic dispersion law. It can be proven by a differentiation of the expression \eqref{onedim_int} with respect to $\mu$ and integration by parts (see also for details \cite{IzvVuz}). Then, under the assumptions supposed in obtaining the above expression for the $\Omega$-potential, we have
\begin{equation}\label{No}
    N^0\approx-\frac{\pi S}{4l_D^2}[\sqrt{\pi}q_m\Li_{3/2}(-e^\mu)+\ln(1+e^\mu)],
\end{equation}
for the main contribution to the average particle number. The oscillating part reads as
\begin{equation}\label{Nos}
    N^{os}\approx\frac{\pi^2S}{l^2_D}\re\left[\sum_{k=0}^{K-1}\ln(1-e^{2\pi iq_mp_k})-\frac{p_K}{2\pi^2 q_m}\Li_2(e^{2\pi iq_mp_K})-\frac{i}{4\pi^3q^2_m}\Li_3(e^{2\pi iq_mp_K})\right].
\end{equation}
This formulas allow us to find a dependence of the chemical potential on the sizes of the crystal.

\begin{figure}[t]
\centering
\includegraphics*[trim=0mm 15mm 0mm 15mm,clip,width=7cm]{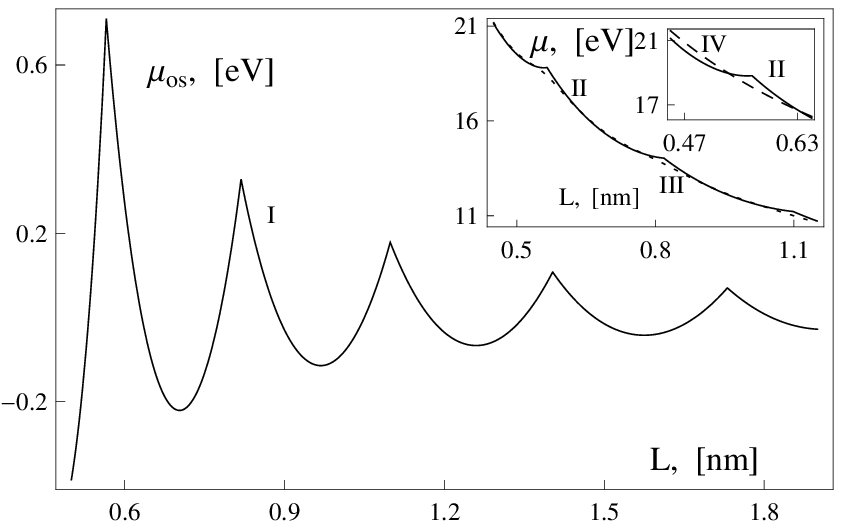}\;%{mu-flate-plate-3.eps}  \;
\includegraphics*[trim=0mm 15mm 0mm 15mm,clip,width=7cm]{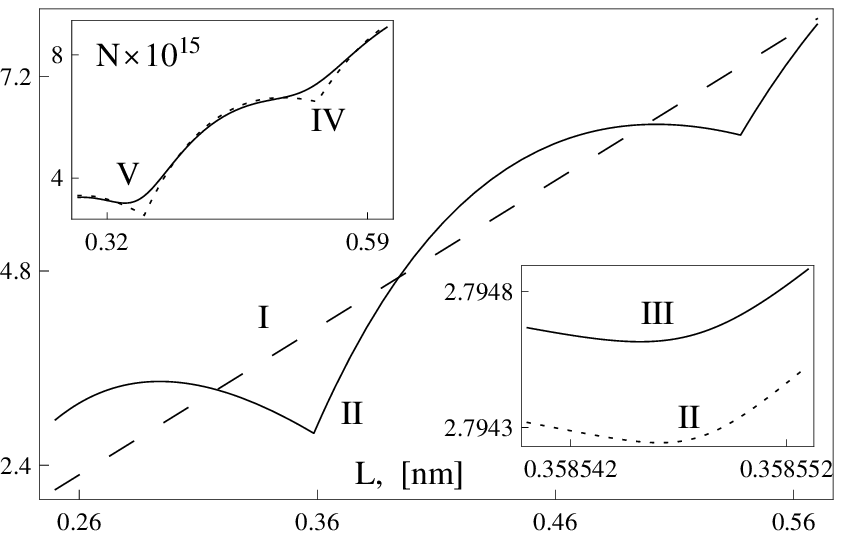}%{n-flate-plate-1.eps}
\caption{{\footnotesize On the left panel: I. The chemical potential of electrons in the thin metal film at the temperature $\beta^{-1}=1$ K and the average particle number $N=1.6\times10^{16}$ what corresponds to the undeformed metal film with the area $S=1$ cm${}^2$, the width $L_x=1$ nm, and the chemical potential $\mu=11.7$ eV. The small plots depict the total chemical potential at $\beta^{-1}=10^{-3}$ K (II) and $\beta^{-1}=10^3$ K (III). The small plot (IV) depicts the quasiclassical part $\tilde{\mu}^0$ of the chemical potential at $\beta^{-1}=1$ K. On the right panel: I. The quasiclassical contribution to the average number of conduction electrons at $\beta^{-1}=1$ K and $\mu=11.7$ eV. The total number of conduction electrons in the thin metal film with the area $S=1$ cm${}^2$ at the fixed chemical potential $\mu=11.7$ eV, and the temperatures $\beta^{-1}=1$ K (II), $\beta^{-1}=1.5$ K (III), $\beta^{-1}=10^{-3}$ K (IV), and $\beta^{-1}=10^3$ K (V).}}\label{plot metal film}
\end{figure}

It is not difficult to express $\mu=\mu(\beta, N,L)$ from Eqs. \eqref{No} and \eqref{Nos} numerically. Though, it is impossible to represent the expression for $\mu(\beta, N,L)$ in terms of known special functions. Therefore, we derive here an approximate expression for the chemical potential making certain assumptions. First, we suppose that the oscillating part is much lesser than the quasiclassical contribution $N^{os}\ll N^0$. Whence
\begin{equation}
    \mu^{os}\approx-\left(\partial_{\mu_0}N^0(\mu_0)\right)^{-1}N^{os}(\mu_0),
\end{equation}
where $\mu^0$ is the solution to Eq. \eqref{No} with respect to $\mu$. In order to obtain an explicit expression for $\mu^0$, we also assume that $\mu\gg1$. In that case, recalling the asymptotic expansion of a polylogarithm (see, e.g., \cite{BatErde,IzvVuz,Wood}), we arrive at
\begin{equation}
    N^0\approx\frac{\pi S}{3l_D^2}\mu\left(q_m\mu^{1/2}-\frac34\right).
\end{equation}
An exact solution to this equation is readily found. However, we give only the approximate value of the dimensionless chemical potential
\begin{equation}
    \mu^0\approx\frac{9}{4q_m^2}\left[\left(\frac{8NL_x^2}{9\pi S}\right)^{1/3}+\frac16\right]^2,
\end{equation}
which is valid when the expression in parenthesis is much greater than unity, i.e., when the film thickness $L_x$ is not too much small. Making use of this expression, we have under the same assumptions
\begin{equation}
    \partial_{\mu_0} N^0\approx\frac{3\pi S}{4l_D^2}\left[\left(\frac{8NL_x^2}{9\pi S}\right)^{1/3}-\frac16\right],
\end{equation}
what allows us to obtain the oscillating part of the chemical potential. Now it is a rather simple task to find the period of oscillations at a fixed number of the conduction electrons and variable chemical potential
\begin{equation}\label{period}
    \ell\approx\left(\frac{9\pi S L_x}{8N}\right)^{1/3}.
\end{equation}
The oscillation period increases with the film thickness. In terms of the oscillation period \eqref{period}, the above used assumption on the film thickness turns into the condition $L_x\gg\ell$.

In order to estimate the oscillation amplitude of the chemical potential, we observe that an absolute value of the expression \eqref{Nos} possesses a maximum where the argument of the polylogarithms entering this expression is close to unity. Despite the fact that the first term in \eqref{Nos} has a logarithmic singularity at this point and the rest two are finite in unity, the main contribution for the conduction electrons in metals comes from the second term in Eq. \eqref{Nos}. Under the conditions considered by us, this term is approximately $10^3$ times larger than the first term. The third term is much smaller than the second one and gives a vanishing contribution when the argument of the polylogarithm is equal to unity. Keeping in mind that the polylogarithm in unity is expressed through the $\zeta$-function \eqref{polylog def}, we find the following estimation for the oscillation amplitude of the dimensionless chemical potential:
\begin{equation}\label{amplitude}
    |\mu^{os}|\lesssim\frac{\pi}{6q_m^2},
\end{equation}
Thus we see that the oscillation amplitude decreases quadratically with increasing of the film thickness. The obtained estimations for the period and amplitude of oscillations are in a good agreement with the numerical results presented on Fig. \ref{plot metal film}.

\begin{acknowledgments}

The work is supported by the Russian Ministry of Education and Science, contract No 02.740.11.0238, the FTP ``Research and Pedagogical Cadre for Innovative Russia'', contracts No P1337, P2596, and the RFBR grant 09-02-00723-a.

\end{acknowledgments}

%\newpage


\begin{thebibliography}{99}

\selectlanguage{english}

%(1)-too old
\bibitem{VasilHeatKer}
D.~V. Vassilevich,
\textsl{Heat kernel expansion: user's manual},
Phys. Rep. \textbf{388}, 279 (2003),
hep-th/0306138.

%(2)-too old
\bibitem{DowKen}
J.~S. Dowker and G. Kennedy,
\textsl{Finite temperature and boundary effects in static space-times},
J. Phys. A: Math. Gen. \textbf{11}, 895 (1978).

%(3)- too old
\bibitem{KirstHKE}
K. Kirsten,
\textsl{Grand thermodynamic potential in a static spacetime with boundary},
Class. Quantum Grav. \textbf{8}, 2239 (1991).

%(4)- too old
\bibitem{Gilkey}
P.~B. Gilkey,
\textsl{Invariance Theory, the Heat Equation, and the Atiyah-Singer Index Theorem},
(Publish or. Perish, Wilmington, Delaware, 1984).

%(5)-too old
\bibitem{LandOsc}
L. Landau, \textsl{Diamagnetismus der Metalle},
Z. Phys. \textbf{64}, 629 (1930).

%(6)-book
\bibitem{Shoenb}
D. Shoenberg,
\textsl{Magnetic Oscillations in Metals},
(Cambridge University Press, Cambridge, 1984).

%(7) -too old- 2 works
\bibitem{SchubdeHa}
L.~W. Shubnikov, J.~W. de Haas,
Leiden Commun. \textbf{207a} (1930); Proc. Netherlands R. Acad. Sci. \textbf{33} 130, 163 (1930).


%(8) -too old 2 works
\bibitem{deHavAlp}
J.~W. de Haas, P.~M. van Alphen,
\textsl{Note on the dependence of the susceptibility of diamagnetic metal on the field},
Leiden Commun. \textbf{208d} (1930); Proc. Netherlands Roy. Acad. Sci. \textbf{33}, 680, 1106 (1930).



%(9) -too old russian
\bibitem{Kulik}
%И.~О. Кулик,
%О размерных осцилляционных эффектах в металлах при произвольном законе дисперсии,
%Письма в ЖЭТФ \textbf{6}, 652 (1967).
I.~O. Kulik , \textsl{Oscillatory size effects in metals with arbitrary dispersion},
Pis'ma Zh. Eksp. Teor. Fiz., \textbf{6}, 652 (1967) [JETP Letters \textbf{6}, 652, (1967)].

%
%%%(10) -too old russian
\bibitem{LifshKos}
%И.~М. Лифшиц, А.~М. Косевич,
%К теории магнитной восприимчивости тонких слоев металлов при низких температурах,
%ДАН  СССР \textbf{92}, 795 (1953)
I.~M. Lifshits and A.~M. Kosevich,
Dokl. Akad. Nauk SSSR, \textbf{92}, 795 (1953);
%К теории эффекта де Гааза-ван Альфена для частиц с произвольным законом дисперсии,
%ДАН  СССР \textbf{96}, 963 (1954)
\textbf{96}, 963 (1954);
%К теории магнитной восприимчивости металлов при низких температурах,
%ЖЭТФ \textbf{29}, 730 (1955)
Zh. Eksp. Teor. Fiz. \textbf{29}, 730 (1955) [Sov.
Phys. JETP \textbf{2}, 636 (1956)];
%Эффект де Гааза-ван Альфена в тонких слоях металлов,
%ЖЭТФ \textbf{29}, 743 (1955)
\textbf{29}, 743 (1955).

%[Sov. Phys. JETP \textbf{2}, 636 (1956)].

%(11) -old
\bibitem{LifshKag}
%И.~М. Лифшиц, М.~И. Каганов,
I.~M. Lifshits, M.~I. Kaganov, \textsl{Some problems of the electron theory of metals}:
%I. Классическая и квантовая механика электронов в металлах,
%УФН \textbf{69}, 419 (1959);
\textsl{I. Classical and quantum mechanics of electrons in metals},
Usp. Fiz. Nauk \textbf{69}, 419 (1959)
[Sov. Phys. Usp. \textbf{2}, 831 (1960)];
%II. Статистическая механика и термодинамика электронов в металлах,
%УФН \textbf{78}, 411 (1962);
\textsl{II. Statistical mechanics and thermodynamics of electrons in metals},
Usp. Fiz. Nauk \textbf{78}, 411 (1962) [Sov. Phys. Usp. \textbf{5}, 878 (1963)];
%III. Кинетические свойства электронов в металле,
% УФН \textbf{87}, 389 (1965)
\textsl{III. Kinetic properties of electrons in metals}, Usp. Fiz. Nauk \textbf{87}, 389 (1965) [Sov. Phys. Usp. \textbf{8}, 805 (1966)];
I.~M. Lifshits, M.~Ya. Azbel and M.~I. Kaganov, \textsl{Electron Theory of Metals} (New York, Consultants Bureau, 1973).

%(12) - too old russian
\bibitem{Nedore}
S.~S. Nedorezov,
\textsl{Effects of boundaries on the thermodynamic properties of a Fermi gas},
Izv. Vyssh. Uchebn. Zaved., Fiz. \textbf{3}, 11 (1965); \textsl{Surface effects in the thermodynamics of conductivity electrons},
Zh. Eksp. Teor. Fiz. \textbf{51}, 1587 (1966); \textsl{Oscillations of the electron thermodynamic characteristics of a metal film at high pressures},
Zh. Eksp. Teor. Fiz. \textbf{51}, 1575 (1966).

%(13) -too old russian
\bibitem{Azbel}
%М.~ Я.~ Азбель,
%К вопросу о восстановлении формы Ферми-поверхности в металлах,
%ЖЭТФ \textbf{34}, 754 (1958).
M.~Ya. Azbel, Zh. Eksp. Teor. Fiz. \textbf{34}, 754 (1958).

%(14) - too old title
\bibitem{OnsagOsc}
%L.~ Onsager,
%\textbf{!!!!!!!!!!!!!!!!!!!!!!!!!!!!!!!!!!!!!!!!!!!1}
%Phil. Mag. 43, 1006 (1952).
L. Onsager,
\textsl{Interpretation of the de Haas-Van Alphen Effect},
Phil. Mag. \textbf{43}, 1006 (1952).

%(15) -too old
\bibitem{CasPold}

H.~B.~G. Casimir,
\textsl{On the attraction between two perfectly conducting plates},
Proc. Kon. Nederland. Akad. Wetensch. B \textbf{51}, 793 (1948);
H.~B.~G. Casimir and D. Polder,
\textsl{The Influence of Retardation on the London-van der Waals Forces}, Phys. Rev. \textbf{73}, 360 (1948).

%(16) -very too old
\bibitem{Epstein}
P. Epstein,
\textsl{Zur Theorie allgemeiner Zetafunktionen},
Math. Ann. \textbf{56}, 615 (1903).

%(17) -too old
\bibitem{ChowSel}
S. Chowla and A. Selberg,
\textsl{On Epsteins zeta-function (I)},
Proc. Natl Acad. Sci. USA \textbf{35}, 371 (1949).

%(18) -old
\bibitem{AmbWolf}
J. Ambj{\o}rn  and S. Wolfram,
\textsl{Properties of the vacuum. I. Mechanical and thermodynamic},
Ann. Phys. \textbf{147}, 1 (1983).

%(19) -arxiv
\bibitem{KirstBo}
K. Kirsten,
\textsl{Spectral Functions in Mathematics and Physics}
(CRC Press, Boca Raton, 2002).

%(20)-arxiv
\bibitem{HJKS}
M.~P. Hertzberg, R.~L. Jaffe, M. Kardar, and A. Scardicchio,
\textsl{Casimir forces in a piston geometry at zero and finite temperatures},
Phys. Rev. D \textbf{76}, 045016 (2007),
arXiv:0808.0047.

%%(21)-arxiv
\bibitem{LimTeo}
S.~C. Lim and L.~P. Teo,
\textsl{Finite temperature Casimir energy in closed rectangular cavities: a rigorous derivation based on a zeta function technique},
J. Phys. A: Math. Theor. \textbf{40}, 11645 (2007),
arXiv:0804.3916.


%(22)-arxiv
\bibitem{ElOdSah}
E. Elizalde, S.~D. Odintsov, and A.~A. Saharian,
\textsl{Repulsive Casimir effect from extra dimensions and Robin boundary conditions: From branes to pistons},
Phys. Rev. D \textbf{79}, 065023 (2009),
arXiv:0902.0717
	
%(23)-arxiv
\bibitem{Edery}
A. Edery,
\textsl{Multidimensional cut-off technique, odd-dimensional Epstein zeta functions and Casimir energy of massless scalar fields},
J. Phys. A: Math. Gen. \textbf{39}, 685 (2006); \textsl{Casimir piston for massless scalar fields in three dimensions},
Phys. Rev. D \textbf{75}, 105012 (2007); A. Edery, V.~N. Marachevsky,
\textsl{Compact dimensions and the Casimir effect: the Proca connection},
JHEP \textbf{12}, 035 (2008).

%(24)-arxiv
\bibitem{Berndt}
B.~C. Berndt,
\textsl{Identities involving the coefficients of a class of Dirichlet series. VI},
Trans. Am. Math. Soc. \textbf{160}, 157 (1971).

%(25)-arxiv
\bibitem{KTTY}
Sh. Kanemitsu, Y. Tanigawa, H. Tsukada and M. Yoshimoto,
\textsl{On Bessel series expressions for some lattice sums: II},
J. Phys. A: Math. Gen. \textbf{37}, 719 (2004).

%(26)-
\bibitem{Elizalde}
E. Elizalde,
\textsl{Zeta function methods and quantum fluctuations},
J. Phys. A: Math. Theor. \textbf{41}, 304040 (2008),
arXiv:0712.1346.

%(27)-old
\bibitem{polylog}
H.~E. Haber, H.~A. Weldon,
\textsl{On the relativistic Bose-Einstein integrals},
J. Math. Phys. \textbf{23}, 1852 (1981);
\textsl{Finite-temperature  symmetry breaking and Bose-Einstein condensation},
Phys. Rev. D \textbf{25}, 502 (1982).

%(28)-too old russian
\bibitem{Rumer}
%Ю.~Б.~ Румер,
%К теории магнетизма электронного газа,
%ЖЭТФ \textbf{18}, 1081 (1948).
%\texbf{!!!!!!!!!!!!!!!!!!!!!!!!!!!!!!!!!!!!!!!!!!!!!!!!!!!!!!}
Yu.~B. Rumer, Zh. Eksp. Teor. Fiz. \textbf{18}, 1081 (1948).


%(29) -too old russian
\bibitem{Zilb}
%Г.~Е.~ Зильберман,
%Магнитные свойства металлов при низких температурах,
%ЖЭТФ \textbf{21}, 1209 (1951).
%\texbf{!!!!!!!!!!!!!!!!!!!!!!!!!!!!!!!!!!!!!!!!!!!!!!!!!!!!!!}
G.~E. Zilberman, Zh. Eksp. Teor. Fiz. \textbf{21}, 1209 (1951).


%(30)-too old russian
\bibitem{BregZhuch}
%А.~Х. Брегер, А.~А. Жуховицкий,
%Поверхностное натяжение металлов,
%ЖФХ \textbf{20}, 355 (1946).
A.~Kh. Breger and A.~A. Zhukhovitskii, Zh. Fiz. Khim. \textbf{14}, 569 (1946).

%(31)-arxiv
\bibitem{GJPVW}
V.~M. Gvozdikov, A.~G.~M. Jansen, D.~A. Pesin, I.~D. Vagner, and P. Wyder,
\textsl{Quantum magnetic oscillations of the chemical potential in superlattices and layered conductors},
Phys. Rev. B \textbf{68}, 155107 (2003);
\textsl{de Haas-van Alphen and chemical potential oscillations in the magnetic-breakdown quasi-two-dimensional organic conductor} $\kappa$-(BEDT-TTF)${}_2$Cu(NCS)${}_2$,
Phys. Rev. B \textbf{70}, 245114 (2004).

%(32) -too old
\bibitem{Wallace}
R.~P. Wallace,
\textsl{The band theory of graphite},
Phys. Rev. \textbf{71}, 622 (1946).

%(33)-too old
\bibitem{McCl}
J.~W. McClure,
\textsl{Diamagnetism of graphite},
Phys. Rev. \textbf{104}, 666 (1956).

%(34)-arxiv
\bibitem{Mostep}
G.~L. Klimchitskaya, U. Mohideen, and V.~M. Mostepanenko,
\textsl{The Casimir force between real materials: Experiment and theory},
Rev. Mod. Phys. \textbf{81}, 1827 (2009),
arXiv:0902.4022.

%(35)- works
\bibitem{KatNov}
M.~I. Katsnelson and K.~S. Novoselov,
\textsl{Graphene: new bridge between condensed matter physics and quantum electrodynamics},
Solid State Commun. \textbf{143}, 3 (2007),
arXiv:cond-mat/0703374.

%(36)-arxiv
\bibitem{NGMJKGDF}
K.~S. Novoselov, A.~K. Geim, S.~V. Morozov, D. Jiang, M.~I. Katsnelson, I.~V. Grigorieva, S.~V. Dubonos and A.~A. Firsov,
\textsl{Two-dimensional gas of massless Dirac fermions in graphene},
Nature \textbf{438}, 197 (2005),
arXiv:cond-mat/0509330.

%(37)-
\bibitem{NGPNG}
A.~H. Castro Neto, F. Guinea, N.~M.~R. Peres, K.~S. Novoselov and A.~K. Geim,
\textsl{The electronic properties of graphene},
Rev. Mod. Phys. \textbf{81}, 109 (2009),
arXiv:0709.1163.

%(38)-works
\bibitem{FuPiGoMo}
J.~N. Fuchs, F. Pi\'{e}chon, M.~O. Goerbig and G. Montambaux,
\textsl{Topological Berry phase and semiclassical quantization of cyclotron orbits for two dimensional electrons in coupled band models},
Eur. Phys. J. B \textbf{77}, 351 (2010),
arXiv:1006.5632.

%(39)- arxiv
\bibitem{VshiKli}
A.~S. Vshivtsev, K.~G. Klimenko, and B.~V. Magnitskii,
\textsl{Landau oscillations in (2+1)-dimensional quantum electrodynamics},
Zh. Eksp. Theor. Fiz. \textbf{107}, 307 (1995)
[J. Exp. Theor. Phys. \textbf{80}, 162 (1995)]; A.~S. Vshivtsev and K.~G. Klimenko,
\textsl{An exact expression for magnetic oscillations in quantum electrodynamics},
Zh. Eksp. Theor. Fiz. \textbf{109}, 954 (1996)
[J. Exp. Theor. Phys. 82, \textbf{154} (1996)].

%(40) arxiv
\bibitem{CarmUll}
P. Carmier and D. Ullmo,
\textsl{Berry phase in graphene: Semiclassical perspective},
Phys. Rev. B \textbf{77}, 245413 (2008).

%(41) arxiv
\bibitem{SharGus}
S.~G. Sharapov, V.~P. Gusynin, and H. Beck,
\textsl{Magnetic oscillations in planar systems with the Dirac-like spectrum of quasiparticle excitations},
Phys. Rev. B \textbf{69}, 075104 (2004); V.~P. Gusynin, and S.~G. Sharapov,
\textsl{Magnetic oscillations in planar systems with the Dirac-like spectrum of quasiparticle excitations. II. Transport properties},
Phys. Rev. B \textbf{71}, 125124 (2005).

%(42) arxiv/works
\bibitem{BFGV}
M. Bordag, I.~V. Fialkovsky, D.~M. Gitman, and D.~V. Vassilevich,
\textsl{Casimir interaction between a perfect conductor and graphene described by the Dirac model},
Phys. Rev. B \textbf{80}, 245406 (2009)
arXiv:0907.3242; I.~V. Fialkovsky, V.~N. Marachevsky, and D.~V. Vassilevich,
\textsl{Finite temperature Casimir effect for graphene}, arXiv:1102.1757.

%(43) arxiv
\bibitem{GuLiZho}
G. Gui, J. Li, and J. Zhong,
\textsl{Band structure engineering of graphene by strain: First-principles calculations},
Phys. Rev. B \textbf{78}, 075435 (2008).

%(43-1)
\bibitem{SoCoLo} Y.-W. Son, M.~L. Cohen, and S.~G. Louie, \textsl{Energy gaps in graphene nanoribbons},
Phys. Rev. Lett. \textbf{97}, 216803 (2006).

%(44) arxiv
\bibitem{PGNLCH}
J.~E. Proctor, Eu. Gregoryanz, K.~S. Novoselov, M. Lotya, J.~N. Coleman, and M.~P. Halsall,
\textsl{Graphene under hydrostatic pressure},
Phys. Rev. B \textbf{80}, 073408 (2009),
arXiv:0905.3103.

%(45) book russian
\bibitem{Wats}
%Ватсон  Г.~Н.
%Теория бесселевых функций. т. 1.
%-- М.: Издательство иностранной литературы, 1949. -- 799 с.
G.~N. Watson,
\textsl{A Treatise on the Theory of Bessel Functions}
(CUP, Cambridge, 1944).
%(T)(814s).

%(46) work-
\bibitem{Sahar}
A.~A. Saharian,
\textsl{The generalized Abel-Plana formula with applications to Bessel functions and Casimir effect},
arXiv:0708.1187.

%(47) book russian
\bibitem{GrRy}
%Градштейн И.С., Рыжик И.М.
%Таблицы интегралов, сумм, рядов и произведений.
%-- М.: Физматлит, 1962. -- 1100 с.
I.~S. Gradshteyn, I.~M. Ryzhik,
\textsl{Table of Integrals, Series, and Products}
(Acad. Press, Boston, 1994).
%1632-p
%ISBN: 978-0122947551, 012294755X

%(48) book russian
\bibitem{LandLifshStat}
%Ландау Л.~Д., Лифшиц Е.~М.
%Статистическая физика. ч. 1.
%-- М.: Физматлит, 2002. -- 616 с.
L.~D. Landau, E.~M. Lifshitz,
\textsl{Statistical Physics. Part I}
(Pergamon, Oxford, 1978).

%(49) -
\bibitem{Fedoryuk}
M.~V. Fedoryuk,
\textsl{The Method of Steepest Descent}
(Nauka, Moscow, 1977) [in Russian].

%(50) book russian
\bibitem{BatErde}
%Г. Бейтмен, А. Эрдейи,
%Высшие трансцендентные функции. т. 1.
%-- М.: Наука, 1973. -- 296 с.
%H.~ Bateman, A.~ Erdelyi,
%Higher transcendental functions. Vol. 1
%(Nauka, Moscow, 1953)

H. Bateman, A. Erdelyi,
\textsl{Higher Transcendental Functions Vol. 1}
(McGraw-Hill, New York, 1953).

%(51) book russian
\bibitem{PrBrMaIII}
%Прудников А.~П., Брычков Ю.~А., Маричев О.~И.
%Интегралы и ряды. т. 3: Специальные функции. Дополнительные главы.
%-- М.: Физматлит, 2003. -- 688 с.
A.~P. Prudnikov, Yu.~A. Brychkov, and O.~I. Marichev,
\textsl{Integrals and Series, Vol. 3, More Special Functions}
(Gordon \& Breach Sci. Publ., New York, 1990)

%(52)-arxiv
\bibitem{FKKLM}
S.~A. Fulling, L. Kaplan, K. Kirsten, Z.~H. Liu, and K.~A. Milton,
\textsl{Vacuum stress and closed paths in rectangles, pistons and pistols},
J. Phys. A: Math. Theor. \textbf{42}, 155402 (2009),
arXiv:0806.2468.

%(53)- arxiv-no
\bibitem{Marach}
V.~N. Marachevsky,
\textsl{Casimir interaction: pistons and cavity},
J. Phys. A: Math. Theor. \textbf{41}, 164007 (2008).

%(54) -arxiv
\bibitem{LimTeo1}
S.~C. Lim and L.~P. Teo, \textsl{Topological symmetry breaking of self-interacting fractional Klein-Gordon field theories on toroidal spacetime},
J. Phys. A: Math. Theor. \textbf{41}, 145403 (2008), arXiv:0804.3910; \textsl{Repulsive Casimir force at zero and finite temperature},
New J. Phys. \textbf{11}, 013055 (2009), arXiv:0812.0426; \textsl{Finite-temperature Casimir effect in piston geometry and its classical limit},
Eur. Phys. J. C  \textbf{60}, 323 (2009); \textsl{Repulsive Casimir force from fractional Neumann boundary conditions},
Phys. Lett. B \textbf{679}, 130 (2009), arXiv:0906.0635; L.~P. Teo,
\textsl{Finite-temperature Casimir pistons for an electromagnetic field with mixed boundary conditions and its classical limit},
J. Phys. A: Math. Theor. \textbf{42}, 105403 (2009); \textsl{Casimir piston of real materials and its application to multilayer models},
Phys. Rev. A \textbf{81}, 032502 (2010).

%(55) -old
\bibitem{CNSS}
F. Caruso, N.~P. Neto, B.~F. Svaiter, N.~F. Svaiter,
\textsl{Attractive or repulsive nature of Casimir force in $D$-dimensional Minkowski space-time},
Phys. Rev. D \textbf{43}, 1300 (1991).

%(56)-old
\bibitem{LCLZ}
Xin-zhou Li, Hong-bo Cheng, Jie-ming Li, and Xiang-hua Zhai,
\textsl{Attractive or repulsive nature of the Casimir force for rectangular cavity},
Phys. Rev. D \textbf{56}, 2155 (1997).

%(57) -arxiv-no
\bibitem{JVH}
R. J\'{a}uregui, C. Villarreal, S. Hacyan,
\textsl{Finite temperature corrections to the Casimir effect in rectangular cavities with perfectly conducting walls},
Ann. Phys. (NY) \textbf{321}, 2156 (2006).

%(58) work-preprint CERN
\bibitem{MREGJK}
M.~F. Maghrebi, S.~J. Rahi, Th. Emig, N. Graham, R.~L. Jaffe, and M. Kardar,
\textsl{Casimir force between sharp-shaped conductors},
arXiv:1010.3223.

%(59) arxiv
\bibitem{GeKlMo}
B. Geyer, G.~L. Klimchitskaya, V.~M. Mostepanenko,
\textsl{Thermal Casimir effect in ideal metal rectangular boxes},
Eur. Phys. J. C  \textbf{57}, 823 (2008),
arXiv:0808.3754.

%(60) arxiv
\bibitem{BreFer}
L. Brey and H.~A. Fertig,
\textsl{Electronic states of graphene nanoribbons studied with the Dirac equation},
Phys. Rev. B \textbf{73}, 235411 (2006),
arXiv:0802.1385.

%(61) old
\bibitem{NakFuj}
K. Nakada and M. Fujita,
\textsl{Edge state in graphene ribbons: Nanometer size effect and edge shape dependence},
Phys. Rev. B \textbf{54}, 17954 (1996).

%(62) -too old-
\bibitem{Pieirls}
R. Peierls, \textsl{Zur Theorie des Diamagnetismus von Leitungselektronen},
Z. Phys. \textbf{80}, 763 (1933); \textsl{Zur Theorie des Diamagnetismus von Leitungselektronen. II. Starke Magnetfelder},
Z. Phys. \textbf{81}, 186 (1933).

%(63) -?
\bibitem{IzvVuz}
P.~O. Kazinski and M.~A. Shipulya,
\textsl{Nonextensive corrections to the one-loop omega-potential for systems with quadratic dispersion law},
Russ. Phys. J. (2011) (to be published).

%(64)
\bibitem{Wood}
D. Wood,
\textsl{The Computation of Polylogarithms, Technical Report 15-92},
Canterbury, UK: University of Kent Computing Laboratory.
http://www.cs.kent.ac.uk/pubs/1992/110.





\end{thebibliography}
\end{document}